\documentclass[ twocolumn,08 pt, amsmath,amssymb, aps,fleqn]{revtex4-1}
\usepackage[hidelinks]{hyperref}
\usepackage[english]{babel}
\usepackage{mathtools}
\usepackage{tabularx}
\usepackage{multirow}
\usepackage{comment}
\usepackage{graphicx}
\usepackage{color}
\usepackage{soul}
\begin{document}
\preprint{APS/123-QED}
\title{The one step fermionic ladder}
\author{Joy Prakash Das} \author{Girish S. Setlur}\email{gsetlur@iitg.ernet.in}
\affiliation{Department of Physics \\ Indian Institute of Technology  Guwahati \\ Guwahati, Assam 781039, India}
\begin{abstract}
The one step fermionic ladder refers to two parallel Luttinger Liquids (poles of the ladder) placed such that there is a finite probability of electrons hopping between the two poles at a pair of opposing points along each of the poles. The many-body Green function for such a system is calculated in presence of forward scattering interactions using the powerful non-chiral bosonization technique (NCBT). This technique is based on a non-standard harmonic analysis of the rapidly varying parts of the density fields appropriate for the study of strongly inhomogeneous ladder systems. The closed analytical expression for the correlation function obtained from NCBT is nothing but the series involving the RPA (Random Phase Approximation) diagrams in powers of the forward scattering coupling strength resummed to include only the most singular terms with the source of inhomogeneities treated exactly. Finally the correlation functions are used to study physical phenomena such as Friedel oscillations and the conductance of such systems with the potential difference applied across various ends.
\end{abstract}

\maketitle
\thispagestyle{empty}
\section{Introduction}

\begin{figure}[b!]
\centering
\includegraphics[scale=0.35]{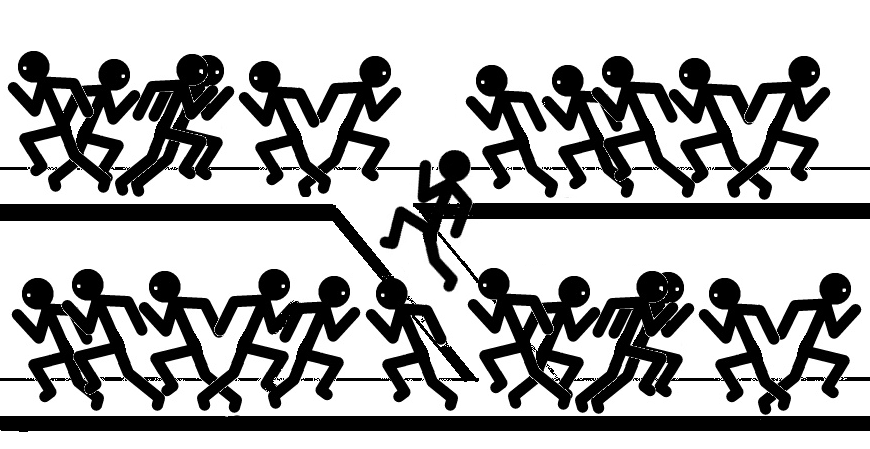}
\caption{ \footnotesize The one step fermion ladder: The two parallel tracks representing the two Luttinger liquids, the athletes representing electrons moving in both the directions, with the fastest athlete possessing the Fermi momentum, while rubbing shoulders against each other representing forward scattering interactions. One athlete running between the tracks represents the hopping of electrons between the Luttinger Liquids. }
\label{caricature}
\end{figure}

\noindent
One dimensional systems occupy a special position when it comes to inter-particle interactions which are totally different from their higher dimensional counterparts, leading to a state described as Luttinger Liquid (LL) \cite{haldane1981Luttinger}. The primary goal of many-body physics is to obtain the ``N-point Green functions" of a system of many mutually interacting particles in the thermodynamic limit. For fermions in one dimension, the well known analytical method to do so goes under the  name `g-ology' (see e.g. Giamarchi \cite{giamarchi2004quantum}) which works well in translationally invariant systems and also in weakly inhomogeneous systems. A non-conventional approach has been recently developed which easily deals with a particular class of strongly inhomogeneous systems, viz. one with a finite number of barriers and wells clustered around an origin \cite{das2016quantum}. In the present work, the same approach, known as the `Non Chiral Bosonization technique' or NCBT, has been employed to obtain the Green functions for the well-studied  one-step fermionic ladder. It is based on a non-standard harmonic analysis of the rapidly varying parts of the density fields appropriate for the study of such strongly inhomogeneous  systems. This method provides analytical expressions for the most singular part of the asymptotic Green functions without having to use renormalization group (RG) or numerical techniques. The system under consideration is described in Fig. \ref{caricature} in the form of a caricature.

The Luttinger Liquid theory \cite{haldane1981Luttinger}  which is based on the linearization of the dispersion relations of the constituent particles finds its applications in a variety of 1D systems, prominent among them are carbon nanotubes \cite{ishii2003direct}, quantum wires \cite{auslaender2002tunneling}, organic conductors \cite{schwartz1998chain}, ultra cold atoms \cite{bloch2008many}, spin ladder sytems \cite{dagotto1999experiments}, etc. A clean Luttinger liquid (without any impurity) behaves as a perfect conductor and the inter-particle interactions in such systems are handled well using conventional bosonization techniques \cite{giamarchi2004quantum}. But the introduction of even a weak impurity can bring about drastic changes in the system which can be as extreme as the `cutting the chain' phenomenon in the case of repulsive mutual interactions, which was first studied in the seminar paper by Kane and Fisher \cite{kane1992transport}. This  phenomenon is seen more directly using the conductance formula obtained from the correlation functions calculated using the non-chiral bosonization technique \cite{das2016quantum}. A variant of this system is the one-step ladder i.e. Luttinger Liquids (two ``poles") lying close to each other with a non-zero hopping probability from one pole to another at a specific location on each pole.

There have been numerous attempts made to compute the correlation functions of fermionic ladder both numerically \cite{noack1994correlations,PhysRevB.52.6796} and analytically \cite{PhysRevB.47.10461,PhysRevB.46.3159,PhysRevB.50.252,PhysRevB.53.12133}.  H. J. Schulz investigated the phase diagram and excitation spectrum of two parallel Luttinger liquids coupled by single-particle hopping \cite{schulz1996phases}. Patrick et.al. applied the Lieb-Schultz-Mattis Theorem to spinful electrons interacting on a ladder which allowed them to obtain a generalized Luttinger theorem for such systems \cite{gagliardini1998generalization}. D.G. Clarke et.al. demonstrated that there is no coherent single particle hopping between two spin-charge separated Luttinger Liquids for hopping parameter below a critical value and in such situations, the two Luttinger liquids will not exhibit split Fermi surfaces.\cite{clarke1994incoherence}.  S. Das et.al. studied the transport of quasiparticles between two edges of Quantum hall liquid via an anti-dot providing the local scattering \cite{das2008quantum}.

The objective of the present work is to obtain the power law behavior of the correlation functions of the one step ladder in the presence of forward scattering interaction among the particles. This happens to be the most singular behavior of the asymptotic forms of the correlation functions under study. This enables various studies including Friedel oscillations in the density correlation functions and the finite temperature d.c. conductance of such systems.

\section{Problem overview}

Consider the one step ladder where we have two Luttinger Liquids placed parallel to each other such that there is a finite probability of hopping at x=0.
\begin{figure}[h!]
\centering
\includegraphics[scale=0.6]{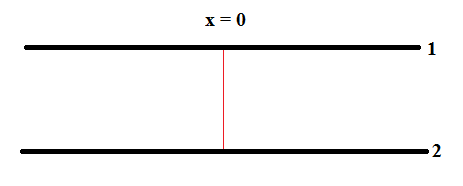}
\caption{\small The one step ladder: Two Luttinger Liquids (1 and 2) placed parallel to each other with a finite probability of hopping at x$=$0.}
\label{onestep}
\end{figure}
The Hamiltonian of the system may be written as follows:\\
\begin{equation}
\begin{aligned}
H=& \sum_k\sum_{j=1,2} \epsilon_k c_{kj}^{\dagger} c_{kj}+\frac{w}{L} \sum_{k,k'} c_{k1}^{\dagger}c_{k'2}+\frac{w}{L} \sum_{k,k'} c_{k'2}^{\dagger}c_{k1}\\
&\hspace{1cm}+ \frac{1}{2} \int^{ \infty}_{-\infty} dx \int^{\infty}_{-\infty} dx^{'} \mbox{  }v(x-x^{'}) \mbox{   } \rho(x) \rho(x^{'})
\end{aligned}
\end{equation}
where $ v(x-x^{'}) = \frac{1}{L} \sum_{q} e^{ -i q(x-x^{'}) } v_q $ (where $ v_q = 0 $ if $ |q| > \Lambda $ for some fixed $ \Lambda \ll k_F $ and $ v_q = v_0 $ is a constant, otherwise) is the forward scattering mutual interaction. `$ L $' is the length of the system and `$ w $' is the hopping parameter which determines the probability of an electron to jump from one pole to another along the $x=0$ line.

The study is carried out in the RPA (Random Phase Approximation) limit, which is a pre-requisite in order to obtain closed expressions of the Green functions, and which means allowing the the Fermi
momentum and the mass of the fermion to diverge in such a way that their ratio is finite (i.e. $ k_F, m \rightarrow \infty $ but $ \frac{ k_F}{m} = v_F < \infty  $)  and thus  linearizing the energy momentum dispersion near the Fermi surface. Units are chosen such that $ \hbar = 1 $, so that $ k_F $ is both the Fermi momentum as well as a wavenumber.

\section{Green's functions of free fermions}
The full two-point Green function (also known as single-particle Green's function) of the system before taking the RPA limit (i.e. with parabolic energy momentum relation) is denoted as \small $
 <T\mbox{  }\psi(x,\sigma,t)\psi^{\dagger}(x',\sigma',t')> $ \normalsize. The time ordering decides whether it is particle or hole Green's function that is being studied and $ \sigma $ is the spin projection of the individual fermions. In terms of this, the asymptotic or RPA Green function is defined by ``smearing out" the positions and times over the scale of the Fermi wavelength and Fermi times as follows,
\small
\begin{equation}
\begin{aligned}
\label{TWOPOINT}
\hspace*{-0.3 cm}
\langle &T\mbox{  }\psi_{\nu}(x,\sigma,t)\psi_{\nu'}^{\dagger}(x',\sigma',t')\rangle
= \lim_{m \rightarrow \infty } \\&\ll \langle T\psi(y,\sigma,\tau)\psi^{\dagger}(y',\sigma',\tau')\rangle e^{-ik_F(\nu y - \nu' y')}e^{ i E_F (\tau-\tau^{'}) } \gg\\
\end{aligned}
\end{equation}
\normalsize
where,
\begin{equation}
\begin{aligned}
&\ll f(t) \gg \mbox{  } = \mbox{  }  \frac{1}{2T_F} \int^{ t+T_F}_{t-T_F} d\tau \mbox{   }f(\tau)\hspace{1 cm}\\
&\ll g(x) \gg \mbox{  } = \mbox{  } \frac{1}{2\lambda_F} \int^{ x+\lambda_F}_{x-\lambda_F} dy \mbox{   }g(y) \
\end{aligned}
\end{equation}
\normalsize
with \footnotesize$ \lambda_F = 2\pi/k_F $ 	\normalsize  and \footnotesize $ T_F =  2\pi/E_F $,   $ k_F = m v_F $ \normalsize and \footnotesize $ E_F =  (1/2) m v_F^2 $
with \footnotesize $ v_F < \infty $  \normalsize being held fixed. Also, here $ \nu,\nu^{'} = \pm 1 $ correspond to the right and left Fermi points.
When mutual interactions between the fermions are absent it is easy to show that the two-point function has the form (at zero temperature in the RPA sense) shown below (here
`i' and `j' in the subscript denote the pole number),
\small
\begin{equation}
\begin{aligned}
< T \mbox{    }\psi_i(&x,t) \psi^{\dagger}_j(x^{'},t^{'}) >_0 \mbox{ }\\
=&e^{ i k_F (x-x^{'}) }  \mbox{  } \frac{i}{ 2 \pi }\mbox{  } \frac{ \delta_{i,j} }{  (x-x^{'})-v_F(t-t^{'})  }\\
+&e^{ -i k_F (x-x^{'}) }  \mbox{  } \frac{i}{ 2 \pi }\mbox{  } \frac{  \delta_{i,j} }{  -(x-x^{'})-v_F(t-t^{'})  }\\
+&\frac{w}{2\pi}  \frac{ ( v_F\delta_{{\bar{i}},j}+i w \mbox{  } \delta_{i,j})}{(v_F^2 +w^2 )}\mbox{     }  \mbox{  }\frac{e^{ -i k_F (|x|+|x^{'}|) }}{   |x|+|x^{'}| + v_F(t-t^{'})   }\\
+& \frac{w}{2\pi} \frac{ ( v_F\delta_{{\bar{i}},j}-i w \mbox{  } \delta_{i,j})}{(v_F^2 +w^2 )}  \mbox{  }\frac{e^{ i k_F (|x|+|x^{'}|) }}{ |x|+|x^{'}| -v_F  (t-t^{'})    }  \\
\label{INPUT1}
\end{aligned}
\end{equation}
\normalsize
Note that in equation (\ref{INPUT1}), the term $[ (\nu x-\nu' x')-v_F(t-t')] $ (where $\nu = \pm 1$) appears in the denominator. In general, when mutual interactions are incorporated into the Luttinger liquid, this term appears with a non-trivial system dependent exponent viz. as $ [ (\nu x-\nu' x')-v_F(t-t') ]^g $. Listing these $ g $'s and other similar exponents  has been one of the main goals of the NCBT technique since $ g = 1 $ is only when mutual interaction between fermions are absent. It is easy to generalize these results to finite temperature since for this a simple replacement viz. $ \frac{1}{X} \rightarrow \frac{\pi}{ \beta v_F} \mbox{  } csch[ \frac{\pi X}{\beta v_F} ] $ is sufficient where e.g. $ X  \equiv (\nu x-\nu' x')-v_F(t-t') $
 and $ \beta $ is inverse temperature.\\

\subsection{ Density density correlation function}
 In the RPA sense, the density $ \rho(x,t) $ may be ``harmonically analysed" as follows.
\begin{equation}\label{INPUT2}
\rho^i(x,t) = \rho^i_s(x,t) + e^{ 2 i k_F x } \mbox{   }\rho^i_f(x,t) +  e^{ - 2 i k_F x } \mbox{   }\rho^{i*}_f(x,t)
\end{equation}
The slowly varying part of the density $ \rho_s $ (the average density is subtracted out, so this is really the deviation) has an auto-correlation function which when mutual interactions are absent, may be written down using Wick's theorem as follows,
\footnotesize
\begin{equation}
\begin{aligned}
\label{INPUT3}
< T \mbox{  } &\rho_s^i(x,t) \rho_s^j(x^{'},t^{'}) >_0  \mbox{  }\\
=& -\frac{ \delta_{i,j} }{ 4\pi^2 }\sum_{\nu=\pm 1}\frac{1}{( (x-x^{'})-\nu v_F(t-t^{'}))^2  }  \theta(xx')\\
-&\frac{ w^2(v_F^2\delta_{\bar{i},j}+w^2 \delta_{i,j} )}{ 4\pi^2 (v_F^2+w^2)^2}\sum_{\nu=\pm 1}\frac{1}{( (x+x^{'})-\nu v_F(t-t^{'}))^2  }\theta(xx')\\
 -& \frac{v_F^2 (v_F^2 \delta_{i,j} + w^2 \delta_{\bar{i},j})}{4\pi^2(v_F^2+w^2)^2}
\sum_{\nu=\pm 1}\frac{ 1 }{  ( (x-x^{'})-\nu v_F(t-t^{'}))^2  }\theta(-xx')
\end{aligned}
\end{equation}
\normalsize

\section{ Bosonized version of the two point Green functions}
The inversion of the defining relation between currents and densities in the standard bosonization scheme that goes by the name g-ology (see the book by Giamarchi \cite{giamarchi2004quantum}) yields the following relation between $ \psi_{\nu}(x,\sigma,t) $ (where $ \nu = R(+1) \mbox{  }or\mbox{   } L(-1) )$ and the slowly varying part of the density (this is a mnemonic for generating the N-point functions),
\begin{equation}
\begin{aligned}
\psi_{\nu}(x,\sigma,t) \sim e^{ i \theta_{\nu}(x,\sigma,t) }
\label{PSINU}
\end{aligned}
\end{equation}
with the local phase given by the formula,
\small
\begin{equation}
\begin{aligned}
\theta_{\nu}(x,\sigma,t) = \pi \int^{x}_{sgn(x)\infty}& dy \bigg( \nu  \mbox{  } \rho_s(y,\sigma,t)\\
 - & \int^{y}_{sgn(y)\infty} dy^{'} \mbox{ }\partial_{v_F t }  \mbox{ }\rho_s(y^{'},\sigma,t) \bigg)
\end{aligned}
\end{equation}\normalsize
\noindent It  has been argued \cite{das2016quantum} that the prescription in equation (\ref{PSINU}) is merely a mnemonic valid only for nearly homogeneous systems and must not be thought of as an operator identity and should not be used to generate Hamiltonians of systems with strongly inhomogeneous external potentials. In order to validate the computation of N-point function for a cluster of impurities in a Luttinger Liquid, it is necessary to slightly modify the above prescription as follows \cite{das2016quantum},
\begin{equation}
\begin{aligned}
\psi_{\nu_i}(x_i&,\sigma_i,t_i) \rightarrow  \sum_{\gamma_i = \pm 1}\sum_{ \lambda_i \in \{0,1\} }C_{\lambda_{i}  ,\nu_i,\gamma_i}(\sigma_i)\mbox{ }
\theta(\gamma_i x_i) \mbox{ }\\
&e^{ i \theta_{\nu_i }(x_i,\sigma_i,t_i) + 2 \pi i \nu_i \lambda_{i}  \int^{x_i}_{sgn(x_i)\infty} \rho_s(-y_i,\sigma_i,t_i) \mbox{  }dy_i }
\label{PSII}
\end{aligned}
\end{equation}
For two Luttinger liquids with a finite probability of hopping at one point, the pole index `i' (to which Luttinger liquid the Field operator belongs to) also comes into the picture and the prescription is given as follows:\\
\begin{equation}
\begin{aligned}
\psi^i_{\nu_i}(x_i,&\sigma_i,t_i) \rightarrow e^{-i X_{\sigma_i} } e^{ - i X^i } \sum_{\gamma_i = \pm 1}\sum_{ \lambda_i \in \{0,1\} }C^i_{\lambda_{i}  ,\nu_i,\gamma_i}(\sigma_i)\mbox{ }\\
&\theta(\gamma_i x_i) \mbox{ }\mbox{ }e^{ i \theta^i_{\nu_i }(x_i,\sigma_i,t_i)}\mbox{ }\mbox{ } e^{ i \pi \sum_{\sigma > \sigma_i } N_{\sigma} }\mbox{ }\mbox{ }e^{ i \pi \sum_{j > i } N^j }\\
&e^{ 2 \pi i \nu_i \lambda_{i}  \int^{x_i}_{sgn(x_i)\infty}\left(\rho^{\bar{i}}_s(y_i,\sigma_i,t_i) + \sum \limits_{i=i,\bar{i}} \rho^i_s(-y_i,\sigma_i,t_i)\right) \mbox{  }dy_i }\\
\label{PSII}
\end{aligned}
\end{equation}
where $\lambda_i=0, 1$ only. Here $ N^i $ is the total number of fermions (all spins combined) on the i-th pole of the ladder and $ N_{\sigma} $ is the total number of fermions with spin projection $ \sigma $ on both the poles combined.
Also $  [X_{\sigma_i}, N_{\sigma_j}] = i\delta_{\sigma_i,\sigma_j} $ and $  [X^i, N^j] = i\delta_{i,j} $, $ [X^i,X^j] = 0 $ and $  [X_{\sigma_i}, X_{\sigma_j}] = 0 $ are canonical conjugates. These additional global quantities ensure that the up spin field anticommutes with the down spin field and
 different poles also anticommute. All  these global quantities commute with the local operators in the exponent. Fortunately as far as practical calculations go, nothing is lost by treating these global objects as c-numbers since doing so enables the correct correlation functions to be reproduced.

The quantities $C^i_{\lambda_{i}  ,\nu_i,\gamma_i}(\sigma_i)$ are c-numbers which involves cut-offs, etc. and are as such   not obtainable using these techniques. They are fixed by a comparison with the non-interacting N-point functions obtained using Fermi algebra. The term $\rho^i_s(-y_i,\sigma_i,t_i)$, which is the signature of NCBT \cite{das2016quantum} ensures that trivial exponents are obtained  when equation (\ref{PSII}) is used to compute the N-point functions in the sense of RPA and in absence of interactions.

The first task is to derive a prescription for choosing the $\lambda$'s which would lead to the N-point functions of the system (without mutual interactions) identical to what is given by Wick's theorem. The next task is to generalise the slow part of the density density correlation functions in equation (\ref{INPUT3}) to include mutual interactions, which when done in the spirit of RPA gives the following results.
 \small
\begin{equation}
\begin{aligned}
<\rho^1_s(x_1,&t_1,\sigma_1)\rho^1_s(x_2,t_2,\sigma_2) > \\
= -&\frac{1}{4\pi^2}\sum_{\nu=\pm 1}\bigg[ \frac{ v_F}{ 2   v_h }  \left( \mbox{    } \frac{1}{ [| x_1-x_2 | -\nu  v_h(t_1-t_2)]^2 }  \right) \\
&\hspace{1.3 cm}+\sigma_1 \sigma_2 \frac{ 1}{ 2   }  \left( \mbox{    } \frac{1}{ [| x_1-x_2 | -\nu  v_F(t_1-t_2)]^2 }   \right)\\
+&\left(\mbox{  }  \frac{v_Fw^2 \mbox{ }\text{sgn}(x_1x_2)}{2v_h(v_Fv_h+w^2)}  -\frac{v_F^2w^2}{2(v_F^4+2v_Fv_hw^2+w^4)}    \right)\times\\
&\hspace{2 cm}\left( \mbox{    } \frac{1}{ [ | x_1|+|x_2 |-\nu v_h(t_1-t_2) ]^2  } \right) \\
+&\sigma_1 \sigma_2 \left(  \frac{w^2 \mbox{ }\text{sgn}(x_1x_2)}{2(w^2+v_F^2)}-  \frac{w^2v_F^2}{2(w^2+v_F^2)^2}\right) \times\\
& \hspace{1.7 cm}\left( \mbox{    } \frac{1}{ [ | x_1|+|x_2 |-\nu v_F(t_1-t_2) ]^2  }  \right)
 \bigg]\\
 \label{DDOSI1}
\end{aligned}
\end{equation}

\begin{equation}
\begin{aligned}
<\rho^1_s(x_1,&t_1,\sigma_1)\rho^2_s(x_2,t_2,\sigma_2) >\\
 =-& \frac{1}{4\pi^2 }\sum_{\nu=\pm 1}\bigg[ \frac{v_F^2w^2}{2(v_F^4+2v_Fv_hw^2+w^4)}\times\\
& \hspace{1.5 cm}\left( \mbox{    } \frac{1}{ [ | x_1|+|x_2 |+v_h(t_1-t_2) ]^2 }\right) \\
+\sigma_1 &\sigma_2 \frac{w^2v_F^2}{2(w^2+v_F^2)^2} \left( \mbox{    } \frac{1}{ [ | x_1|+|x_2 |-\nu v_F(t_1-t_2) ]^2  }
   \right) \bigg] \\
\label{DDOSI2}
\end{aligned}
\end{equation}

\normalsize
\normalsize
Here \scriptsize $ v_h = \sqrt{v_F^2+\frac{2v_F v_0}{\pi}} $  \normalsize is the holon velocity whereas the spinon velocity is just the Fermi velocity since it is the total density that couples to the short range potential: $ v_n = v_F $. The interaction between fermions is the two-body short range forward scattering potential which just means the potential between two particles at $ x $ and $ x^{'} $ is \footnotesize  $ V(x-x^{'}) = \frac{1}{L}\sum_{|q| < \Lambda } e^{ -i q (x-x^{'}) }\mbox{    }  v_0 $, \normalsize   where $ \Lambda $ is held fixed as the RPA limit is taken. Also, $ \langle T \mbox{ }\rho_n(x_1,t_1)\rho_h(x_2,t_2)\rangle \equiv 0 $. Eq.(\ref{DDOSI1}) and Eq.(\ref{DDOSI2}) boil down to the density density correlation functions of a clean Luttinger liquid when the hopping parameter `w' is zero. It is easy to show that Eq.(\ref{DDOSI1}) and Eq.(\ref{DDOSI2}) are the outcomes of a resummation of the most singular parts of the RPA terms in a perturbation series in powers of the mutual interaction $ v_0 $ carried out using Fermi algebra.

The notion of ``the most singular part" of an expression such as the ones shown above may be made sense of in the following manner. Think of these as function of the time difference $ \tau = t_1-t_2 $ (which they are). In the formulas that are encountered while expanding in powers of the coupling, there are going to be terms of the form (e.g.)
\begin{equation}
 \begin{aligned}
\frac{ A \mbox{  }\tau }{ (\tau - a)^2 } + \frac{B}{(\tau-a_1)(\tau-a_2)}
\label{singular}
\end{aligned}
\end{equation}
The first term is regarded as more singular than the second (if $ a_1 \neq a_2 $) since the former is a second order pole whereas the latter when partial fraction expanded are a sum of two first order poles. In the perturbative expansion of the slow part of the density density correlations,
pretending that Wick's theorem applies at the level of the density fluctuations is tantamount to retaining second order poles and discarding poles of a lower order. The same rule applies when deciding what to retain and what to discard in the perturbation expansion of the single-particle Green function.
 The density density correlation functions as mentioned in equations (\ref{DDOSI1}) and (\ref{DDOSI2}) is perturbatively expanded in powers of the interaction parameter $v_0$. The zeroth order term has an exact match with that of the non interacting density density correlation function as mentioned in equation (\ref{INPUT3}) of the main text. The most singular part of the first order term of the density density correlation functions obtained from conventional perturbation theory are as follows:\\
\footnotesize
{\bf Case I : $x_1$ and $x_2$ are on the same pole and same side of the origin\\}
\begin{equation*}
 \begin{aligned}
\delta<&\rho_1(x_1,t_1) \rho_1(x_2,t_2)>
\\	=&\frac{1}{4\pi^3}\bigg(\frac{(t_1-t_	2)}{ (|x_1-x_2| + v_F(t_1-t_2))^3}-\frac{(t_1-t_2)}{ (|x_1-x_2| - v_F(t_1-t_2))^3}\\
&+\frac{w^4}{(w^2+v_F^2)^2}\Big(\frac{(t_1-t_2)}{ (x_1+x_2 +v_F (t_1-t_2))^3}\\
&\hspace{2 cm}-\frac{(t_1-t_2)}{ (x_1+x_2 - v_F(t_1-t_2))^3}\Big)\bigg)\\
\end{aligned}
\end{equation*}
{\bf Case II : $x_1$ and $x_2$ are on same pole and opposite sides of the origin\\}
\begin{equation*}
 \begin{aligned}
\delta<&\rho_1(x_1,t_1) \rho_1(x_2,t_2)>=\frac{v_F^4}{4\pi^3(w^2+v_F^2)^2}
\\&\hspace{0.7cm}\times\bigg(\frac{(t_1-t_	2)}{ (x_1-x_2 +v_F (t_1-t_2))^3}
-\frac{(t_1-t_2)}{ (x_1-x_2 -v_F (t_1-t_2))^3}\bigg)\\
\end{aligned}
\end{equation*}
{\bf Case III : $x_1$ and $x_2$ are on opposite poles and same of the origin\\}
\begin{equation*}
 \begin{aligned}
\delta<&\rho_1(x_1,t_1) \rho_2(x_2,t_2)>=\frac{v_F^2w^2}{4\pi^3(w^2+v_F^2)^2}
\\&\hspace{0.7cm}\times\bigg(\frac{(t_1-t_	2)}{ (x_1+x_2 + v_F(t_1-t_2))^3}
-\frac{(t_1-t_2)}{ (x_1+x_2 - v_F(t_1-t_2))^3}\bigg)\\
\end{aligned}
\end{equation*}
{\bf Case IV : $x_1$ and $x_2$ are on opposite poles and opposite sides of the origin\\}
\begin{equation*}
 \begin{aligned}
\delta<&\rho_1(x_1,t_1) \rho_2(x_2,t_2)>=\frac{v_F^2w^2}{4\pi^3(w^2+v_F^2)^2}
\\&\hspace{0.7cm}\times\bigg(\frac{(t_1-t_	2)}{ (x_1-x_2 +v_F (t_1-t_2))^3}
-\frac{(t_1-t_2)}{ (x_1-x_2 - v_F(t_1-t_2))^3}\bigg)\\
\end{aligned}
\end{equation*}
\normalsize
It is easy to see that these are identical to the corresponding expansions of (\ref{DDOSI1}) and (\ref{DDOSI2}).
\section{ Full two-point Green's function}
The two point (single-particle) Green function, in the sense of RPA, may be written down  using the correspondence in equation (\ref{PSII}). The main focus is on the anomalous exponents which refer to the constants $ g $ that appear in terms of the form $ ( \nu_1 x_1 - \nu_2 x_2 - v_F(t_1-t_2))^g $  that emerge from this calculation. The necessary step here is to formulate  a prescription for deciding which of the $ \lambda_i's $ are zero and which are one and under what circumstances, after which the  $ g $'s may be uniquely pinned down. This prescription follows unambiguously from the condition that an evaluation (of the 2M-point function)  in the Gaussian (and RPA) sense leads to trivial exponents when mutual interactions between fermions are absent.  The coefficients $ C's $ which  depend on the details of the potentials and cutoffs and other such non-universal features are of lesser importance, as is also the case in the conventional approach. The prescription for obtaining the $ \lambda_i's $ are simple. One needs to consider a general 2M-point function and then  mentally pair up one annihilation operator with one creation operator and create M such pairs. This is merely a mental activity since this pairing (Wick's theorem) is not valid when mutual interactions are present.  Consider one such pair and let the two $ \lambda $'s of this pair be  $  (\lambda_m,\lambda_k) $ where $ k > m $. The constraints are as follows:
\footnotesize
\begin{equation}
\lambda_m = \begin{cases} \lambda_k &\mbox{if } (\nu_m,\nu_k) = (\gamma_m,\gamma_k) \mbox{   }\mbox{or} \mbox{   }(\nu_m,\nu_k) = (-\gamma_m,-\gamma_k) \\
1-\lambda_k &\mbox{if } (\nu_m,\nu_k) = (-\gamma_m,\gamma_k) \mbox{   }\mbox{or}\mbox{   }(\nu_m,\nu_k) = (\gamma_m,-\gamma_k)\end{cases}
\label{PRES}
\end{equation}
\normalsize
 This (unique) prescription guarantees the right trivial exponents in the right places when mutual interactions are turned off. The full Green function in presence of interactions are as follows (Notation: \begin{small}$X_i \equiv (x_i,\sigma_i,t_i)$. \end{small} Furthermore the finite temperature versions of the formulas below are obtained by replacing $ Log[Z] $ by $ Log[ \frac{ \beta v_F}{\pi }\mbox{   } Sinh[ \frac{\pi Z}{ \beta v_F} ] ] $ where $ Z \sim  (\nu x_1 - \nu^{'} x_2 ) - v_a (t_1-t_2)  $ and singular cutoffs ubiquitous in this subject are suppressed in this notation for brevity - they have to be understood to be present. ):\\ \mbox{  } \\
\begin{bf} Case I : $x_1$ and $x_2$ on the same side of the origin and on the same pole\end{bf} \\ \scriptsize
\begin{equation}
\begin{aligned}
\Big\langle T&\mbox{  }\psi_{R}(X_1)\psi_{R}^{\dagger}(X_2)\Big\rangle
= \frac{i}{2\pi} e^{\gamma_1 \log{[4x_1x_2]}}\\
& Exp[ -\sum_{ \substack{\nu,\nu^{'} = \pm 1 \\ a = h,n} } Q_{1,1}(\nu,\nu^{'};a) \mbox{  } Log[ (\nu x_1 - \nu^{'} x_2 ) - v_a (t_1-t_2) ] ] \\
\Big\langle T&\mbox{  }\psi_{L}(X_1)\psi_{L}^{\dagger}(X_2)\Big\rangle
= \frac{i}{2\pi} e^{\gamma_1 \log{[4x_1x_2]}}\\
& Exp[ -\sum_{ \substack{\nu,\nu^{'} = \pm 1 \\ a = h,n} } Q_{-1,-1}(\nu,\nu^{'};a) \mbox{  } Log[ (\nu x_1 - \nu^{'} x_2 ) - v_a (t_1-t_2) ] ] \\
\Big\langle T&\mbox{  }\psi_{R}(X_1)\psi_{L}^{\dagger}(X_2)\Big\rangle
= \frac{i}{2\pi} \frac{w^2}{w^2+v_F^2} \times\\
&\frac{1}{2}(e^{\gamma_1\log{[2x_1]}} e^{(3+\gamma_2) \log{[2x_2]}}+e^{(3+\gamma_2) \log{[2x_1]}}e^{\gamma_1\log{[2x_2]}} )\\
& Exp[ -\sum_{ \substack{\nu,\nu^{'} = \pm 1 \\ a = h,n} } Q_{1,-1}(\nu,\nu^{'};a) \mbox{  } Log[ (\nu x_1 - \nu^{'} x_2 ) - v_a (t_1-t_2) ] ] \\
\Big\langle T&\mbox{  }\psi_{L}(X_1)\psi_{R}^{\dagger}(X_2)\Big\rangle
= \frac{i}{2\pi}\frac{w^2}{w^2+v_F^2} \times \\
&\frac{1}{2}(e^{\gamma_1\log{[2x_1]}} e^{(3+\gamma_2) \log{[2x_2]}} +e^{(3+\gamma_2) \log{[2x_1]}}e^{\gamma_1\log{[2x_2]}})\\
& Exp[ -\sum_{ \substack{\nu,\nu^{'} = \pm 1 \\ a = h,n} } Q_{-1,1}(\nu,\nu^{'};a) \mbox{  } Log[ (\nu x_1 - \nu^{'} x_2 ) - v_a (t_1-t_2) ] ] \\
\label{SS1}
\end{aligned}
\end{equation}

\small
The values of the exponents  `Q' are mentioned in Table \ref{SSSP}.

 \begin{bf}Case II : $x_1$ and $x_2$ on the same side of the origin and on opposite poles\end{bf} \\ \scriptsize
 \begin{equation*}
\begin{aligned}
<T\mbox{  }&\psi_R(X_1)\psi_R^{\dagger}(X_2)>\mbox{  } = \mbox{  } 0\\
<T\mbox{  }&\psi_L(X_1)\psi_L^{\dagger}(X_2)>\mbox{  } = \mbox{  }0\\
\end{aligned}
\end{equation*}
 \begin{equation*}
\begin{aligned}
<T\mbox{  }& \psi_R(X_1)\psi_L^{\dagger}(X_2)> \mbox{  } = \mbox{  } \\
&\frac{1}{2\pi^2}\frac{wv_F}{w^2+v_F^2} \frac{e^{\gamma_1\log{[2x_1]}} e^{(3+\gamma_2) \log{[2x_2]}}}{2(x_1-x_2)} \mbox{  } \\ &
Exp[ -\sum_{ \substack{\nu,\nu^{'} = \pm 1 \\ a = h,n} }  S_{1,1}(\nu,\nu^{'};a;1) \mbox{  } \mbox{  } Log[ (\nu x_1 - \nu^{'} x_2 ) - v_a (t_1-t_2) ] ]
 \\ +&\frac{1}{2\pi^2}  \frac{wv_F}{w^2+v_F^2} \frac{  e^{(3+\gamma_2) \log{[2x_1]}}  e^{\gamma_1\log{[2x_2]}}}{2(x_1-x_2)}
\mbox{  }\\&   Exp[ -\sum_{ \substack{\nu,\nu^{'} = \pm 1 \\ a = h,n} }  S_{1,1}(\nu,\nu^{'};a;2) \mbox{  } \mbox{  } Log[ (\nu x_1 - \nu^{'} x_2 ) - v_a (t_1-t_2) ] ]
\\
\end{aligned}
\end{equation*}
 \begin{equation}
\begin{aligned}
<T\mbox{  }&\psi_L(X_1)\psi_R^{\dagger}(X_2)> \mbox{  }= \mbox{  } \\
& \frac{1}{2\pi^2}\frac{wv_F}{w^2+v_F^2} \frac{e^{\gamma_1 \log{[2x_1]}}e^{(3+\gamma_2)\log{[2x_2]}}}{2(x_1-x_2)} \mbox{   } \\ &
Exp[ -\sum_{ \substack{\nu,\nu^{'} = \pm 1 \\ a = h,n} }  S_{-1,-1}(\nu,\nu^{'};a;1) \mbox{  }  Log[ (\nu x_1 - \nu^{'} x_2 ) - v_a (t_1-t_2) ] ]
\\ +&\frac{1}{2\pi^2}\frac{wv_F}{w^2+v_F^2} \frac{e^{(3+\gamma_2)\log{[2x_1]}} e^{\gamma_1 \log{[2x_2]}}}{2(x_1-x_2)}\mbox{  } \\ & Exp[ -\sum_{ \substack{\nu,\nu^{'} = \pm 1 \\ a = h,n} }  S_{-1,-1}(\nu,\nu^{'};a;2)  \mbox{  } Log[ (\nu x_1 - \nu^{'} x_2 ) - v_a (t_1-t_2) ] ]
 \\\\
\label{SS2}
\end{aligned}
\end{equation}

\small
The values of the exponents  `S' are mentioned in Table \ref{SSDP11}.

 \begin{bf}Case III : $x_1$ and $x_2$ on opposite sides of the origin and on the same pole\end{bf} \\ \scriptsize
 \begin{equation}
\begin{aligned}
<T\mbox{  }& \psi_R(X_1)\psi_R^{\dagger}(X_2)> \mbox{  } =\mbox{  } \\&\frac{i}{2\pi^2}\frac{v_F^2}{w^2+v_F^2} \frac{e^{\gamma_1 \log{[2x_1]}}e^{(3+\gamma_2)\log{[2x_2]}}}{2(x_1+x_2)} \mbox{  } \\ &
Exp[ -\sum_{ \substack{\nu,\nu^{'} = \pm 1 \\ a = h,n} }  U_{1,1}(\nu,\nu^{'};a;1) \mbox{  } \mbox{  } Log[ (\nu x_1 - \nu^{'} x_2 ) - v_a (t_1-t_2) ] ]
 \\ +&\frac{i}{2\pi^2}\frac{v_F^2}{w^2+v_F^2} \frac{e^{(3+\gamma_2)\log{[2x_1]}} e^{\gamma_1 \log{[2x_2]}}}{2(x_1+x_2)}
\mbox{  }\\&   Exp[ -\sum_{ \substack{\nu,\nu^{'} = \pm 1 \\ a = h,n} }  U_{1,1}(\nu,\nu^{'};a;2) \mbox{  } \mbox{  } Log[ (\nu x_1 - \nu^{'} x_2 ) - v_a (t_1-t_2) ] ]
\\
<T\mbox{  }&\psi_L(X_1)\psi_L^{\dagger}(X_2)> \mbox{  }=\mbox{  } \\
& \frac{i}{2\pi^2}\frac{v_F^2}{w^2+v_F^2} \frac{e^{\gamma_1 \log{[2x_1]}}e^{(3+\gamma_2)\log{[2x_2]}}}{2(x_1+x_2)} \mbox{   } \\ &
Exp[ -\sum_{ \substack{\nu,\nu^{'} = \pm 1 \\ a = h,n} }  U_{-1,-1}(\nu,\nu^{'};a;1) \mbox{  }  Log[ (\nu x_1 - \nu^{'} x_2 ) - v_a (t_1-t_2) ] ]
\\ +&\frac{i}{2\pi^2}\frac{v_F^2}{w^2+v_F^2} \frac{e^{(3+\gamma_2)\log{[2x_1]}} e^{\gamma_1 \log{[2x_2]}}}{2(x_1+x_2)}\mbox{  } \\ & Exp[ -\sum_{ \substack{\nu,\nu^{'} = \pm 1 \\ a = h,n} }  U_{-1,-1}(\nu,\nu^{'};a;2)  \mbox{  } Log[ (\nu x_1 - \nu^{'} x_2 ) - v_a (t_1-t_2) ] ]
 \\\\
<T\mbox{  }&\psi_R(X_1)\psi_L^{\dagger}(X_2)>\mbox{  } = \mbox{  } 0\\
<T\mbox{  }&\psi_L(X_1)\psi_R^{\dagger}(X_2)>\mbox{  } = \mbox{  }0\\
\label{OS1}
\end{aligned}
\end{equation}

\small
The values of the exponents  `U' are mentioned in Table \ref{OSSP33}.\\

 \begin{bf}Case IV : $x_1$ and $x_2$ on opposite sides of the origin and on different poles\end{bf} \\ \scriptsize
 \begin{equation*}
\begin{aligned}
<T\mbox{  }& \psi_R(X_1)\psi_R^{\dagger}(X_2)> \mbox{  } =  \mbox{  } \\&\frac{1}{2\pi^2}\frac{wv_F}{w^2+v_F^2} \frac{e^{\gamma_1\log{[2x_1]}}e^{(3+\gamma_2)\log{[2x_2]}}}{2(x_1+x_2)} \mbox{  } \\ &
Exp[ -\sum_{ \substack{\nu,\nu^{'} = \pm 1 \\ a = h,n} }  W_{1,1}(\nu,\nu^{'};a;1) \mbox{  } \mbox{  } Log[ (\nu x_1 - \nu^{'} x_2 ) - v_a (t_1-t_2) ] ]
 \\ +&\frac{1}{2\pi^2}\frac{wv_F}{w^2+v_F^2}\frac{e^{(3+\gamma_2)\log{[2x_1]}} e^{\gamma_1\log{[2x_2]}}}{2(x_1+x_2)}
\mbox{  }\\&   Exp[ -\sum_{ \substack{\nu,\nu^{'} = \pm 1 \\ a = h,n} }  W_{1,1}(\nu,\nu^{'};a;2) \mbox{  } \mbox{  } Log[ (\nu x_1 - \nu^{'} x_2 ) - v_a (t_1-t_2) ] ]
\\
\end{aligned}
\end{equation*}
 \begin{equation}
\begin{aligned}
<T\mbox{  }&\psi_L(X_1)\psi_L^{\dagger}(X_2)> \mbox{  }= \mbox{  } \\
& \frac{1}{2\pi^2}\frac{wv_F}{w^2+v_F^2}\frac{e^{\gamma_1 \log{[2x_1]}}e^{(3+\gamma_2)\log{[2x_2]}}}{2(x_1+x_2)} \mbox{   } \\ &
Exp[ -\sum_{ \substack{\nu,\nu^{'} = \pm 1 \\ a = h,n} }  W_{-1,-1}(\nu,\nu^{'};a;1) \mbox{  }  Log[ (\nu x_1 - \nu^{'} x_2 ) - v_a (t_1-t_2) ] ]
\\ +&\frac{1}{2\pi^2}\frac{wv_F}{w^2+v_F^2}\frac{e^{(3+\gamma_2)\log{[2x_1]}} e^{\gamma_1\log{[2x_2]}}}{2(x_1+x_2)}\mbox{  } \\ & Exp[ -\sum_{ \substack{\nu,\nu^{'} = \pm 1 \\ a = h,n} }  W_{-1,-1}(\nu,\nu^{'};a;2)  \mbox{  } Log[ (\nu x_1 - \nu^{'} x_2 ) - v_a (t_1-t_2) ] ]
 \\\\
<T\mbox{  }&\psi_R(X_1)\psi_L^{\dagger}(X_2)>\mbox{  } = \mbox{  } 0\\
<T\mbox{  }&\psi_L(X_1)\psi_R^{\dagger}(X_2)>\mbox{  } = \mbox{  }0\\
\label{OS2}
\end{aligned}
\end{equation}

\small
The values of the exponents  `W' are mentioned in Table \ref{OSDP44}.\\

\begin{table}[h!]
\caption{Luttinger exponents $ Q_{\nu_1,\nu_2}(\nu,\nu^{'};a) $ for $x_1$ and $x_2$ on the same side of the origin and on the same pole. The analytical formulas for the entries in the table are shown in \hyperref[Appendix A]{Appendix A} .\\}
{\begin{tabular}{c  c  c  c  c}
  \hline
   $ Q_{\nu_1,\nu_2}(\nu,\nu^{'};a) $ \mbox{  } & $\substack{ \nu = 1 ;\\\nu^{'} = 1 } $ & $\substack{ \nu = -1 ; \\\nu^{'} = -1 } $ &
   $ \substack{ \nu = 1; \\\nu^{'} = -1  }$ & $\substack{ \nu = -1 ; \\\nu^{'} = 1 } $  \\
   \hline
 $  \substack{\nu_1 = 1,\nu_2 = 1 ;\\a = h } $ & P \hspace{0.6 cm}& \hspace{0.6 cm}Q\hspace{0.6 cm} & \hspace{0.6 cm}X \hspace{0.6 cm}&\hspace{0.6 cm} X \hspace{0.6 cm} \\[4pt]
   $ \substack{ \nu_1 = 1,\nu_2 = 1 ; \\a = n}  $  & $ 0.5 \mbox{  }  $ & 0 & 0 & 0  \\[4pt]
 $ \substack{\nu_1 = -1,\nu_2 = -1 ; \\a = h}  $  & Q & P & X & X  \\[4pt]
  $\substack{ \nu_1 = -1,\nu_2 = -1 ; \\ a = n}  $  & 0 & $ 0.5 \mbox{  }  $  & 0 & 0  \\[4pt]
  $\substack{  \nu_1 = 1,\nu_2 = -1 ;\\a = h}  $  & S & S & Y & Z  \\[4pt]
   $\substack{  \nu_1 = 1,\nu_2 = -1 ;\\ a = n}  $ & 0 & 0 & $ 0.5 \mbox{  }  $ & 0  \\[4pt]
  $\substack{ \nu_1 = -1,\nu_2 = 1 ; \\a = h}  $  & S & S & Z & Y \\[4pt]
  $\substack{\nu_1 = -1,\nu_2 = 1 ; \\a = n}  $ & 0 & 0 & 0 & $ 0.5 \mbox{  }  $  \\[4pt]
  \hline
\end{tabular}}
\label{SSSP}
\end{table}
\begin{figure*}[t!]
\begin{center}
\includegraphics[scale=0.28]{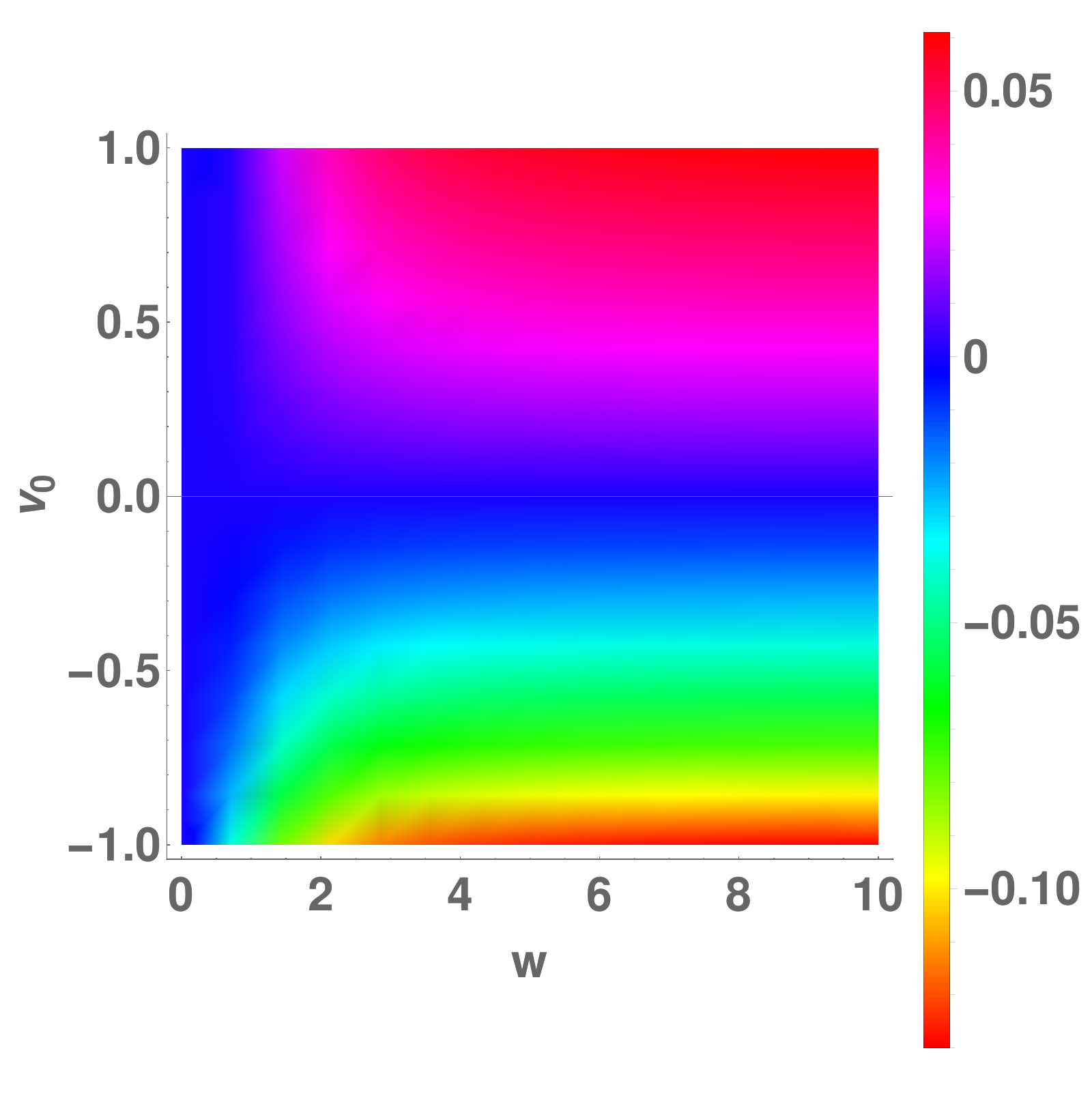}\hspace{0.1 cm}
\includegraphics[scale=0.28]{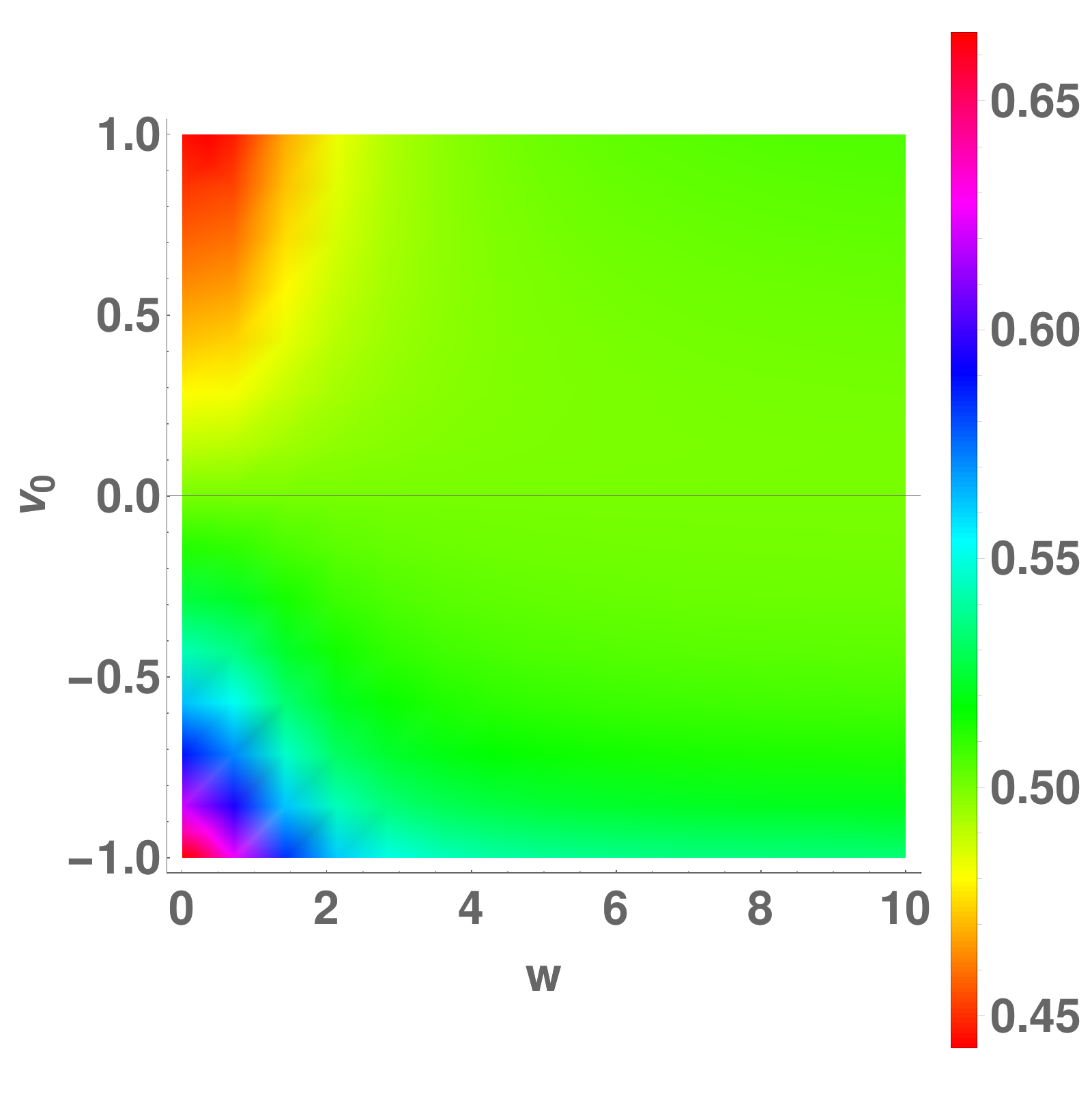}\hspace{0.1 cm}
\includegraphics[scale=0.28]{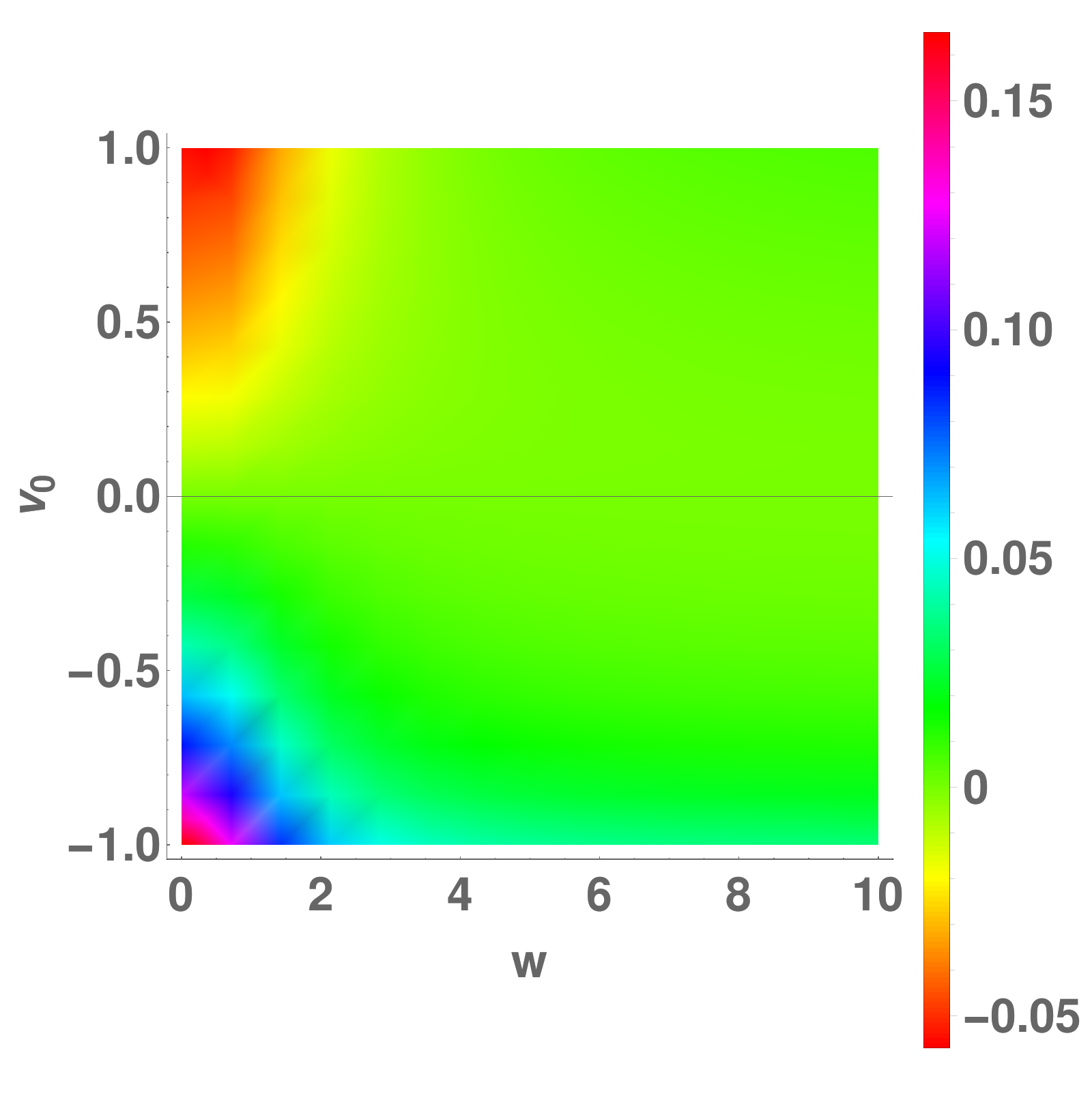}
\scriptsize (a) \hspace{6 cm}(b)\hspace{6 cm} (c)
\end{center}
\caption{ Anomalous exponents vs hopping parameter $w$ and interaction strength $v_0$ for a step ladder ($v_F=1$) with the points on the same pole and same side of the origin (a) X (b) Y (c) Z. The other exponents are independent of the hopping parameter.}
\label{XYZ}
\end{figure*}

\begin{table}[h!]
\caption{Luttinger exponents $ S_{\nu_1,\nu_2}(\nu,\nu^{'};a;j) $ for $x_1$ and $x_2$ on the same side of the origin and on different poles. Explicit expressions for the entries in the table are given in section \hyperref[Appendix A]{Appendix A}.\\}
{ \begin{tabular}{c  c  c  c  c}
 \hline
   $ S_{\nu_1,\nu_1}(\nu,\nu^{'};a;j) $
   \mbox{  }& $\substack{ \nu = 1 ;\\\nu^{'} = 1 } $ & $\substack{\nu = -1 ;\\\nu^{'} = -1 }$ &
   $ \substack{\nu = 1 ; \\\nu^{'} = -1 } $ & $\substack{ \nu = -1 ; \\\nu^{'} = 1 } $  \\
   \hline
 $ \substack{\nu_1 = 1,\nu_2 = 1 ;\\a = h, j = 1  }$ &$A_1$ \hspace{0.6 cm}&\hspace{0.6 cm} $B_1$ \hspace{0.6 cm}&\hspace{0.6 cm} $C_1$\hspace{0.6 cm} &\hspace{0.6 cm} $D_1$\hspace{0.6 cm}  \\[4pt]
   $\substack{ \nu_1 = 1,\nu_2 = 1 ;\\ a = n, j = 1 } $ &-0.5  & $ 0\mbox{  } $  & $ 0.5 \mbox{  }  $ & $ 0 \mbox{  }  $  \\[4pt]
 $\substack{ \nu_1 = 1,\nu_2 = 1 ; \\a = h, j = 2 } $  & $B_1$ & $A_1$ & $C_1$ & $D_1$ \\[4pt]
  $\substack{\nu_1 = 1,\nu_2 = 1 ;\\ a = n, j = 2 }$  & $ 0 \mbox{  } $  &-0.5   & $ 0.5 \mbox{  }  $  & 0  \\[4pt]
  $\substack{ \nu_1 = -1,\nu_2 = -1 ; \\a = h, j = 1}  $  & $A_1$ & $B_1$ & $D_1$ & $C_1$  \\[4pt]
   $\substack{ \nu_1 = -1,\nu_2 = -1 ; \\a = n, j = 1 } $ & -0.5  & $ 0 \mbox{  }  $ & $ 0 \mbox{  }  $ & $ 0.5 \mbox{  }  $  \\[4pt]
  $\substack{ \nu_1 = -1,\nu_2 =- 1 ;\\ a = h, j = 2}  $  & $B_1$ & $A_1$ & $D_1$ & $C_1$\\[4pt]
  $\substack{ \nu_1 = -1,\nu_2 = -1 ;\\ a = n, j = 2 } $ & $0$ & $ -0.5  \mbox{  }  $ & 0 & $ 0.5 \mbox{  }  $  \\[4pt]
  \hline
\end{tabular}}
\label{SSDP11}
\end{table}
\begin{figure}[b!]
\begin{center}
\includegraphics[scale=0.325]{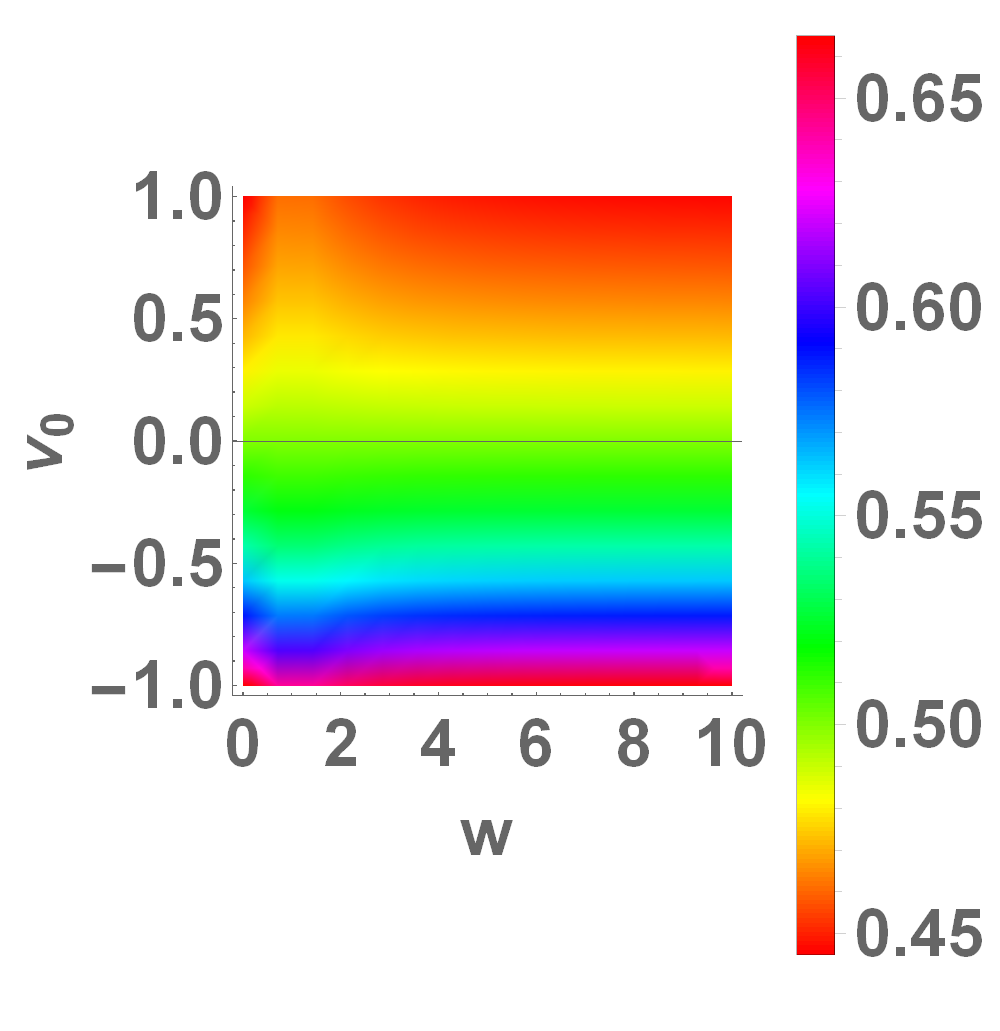}\hspace{0.1 cm}
\includegraphics[scale=0.339]{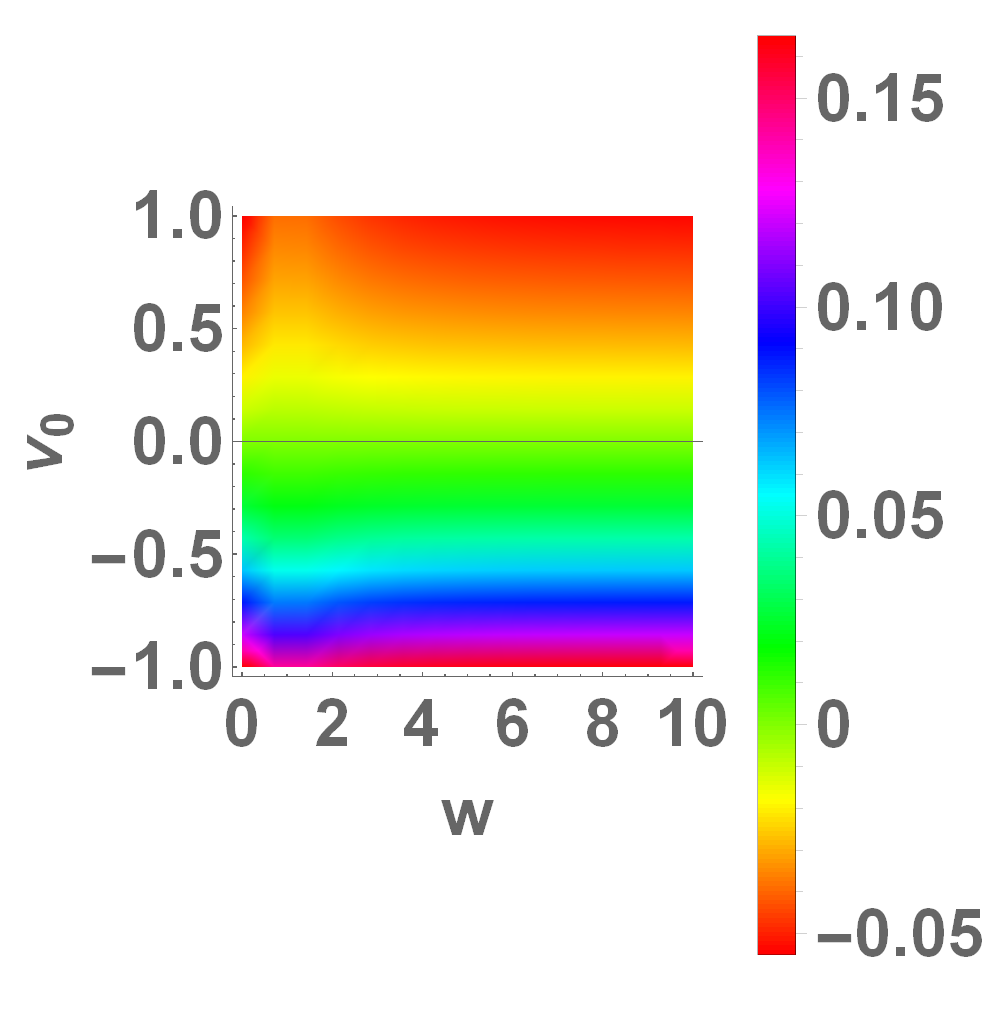}\hspace{0.1 cm}\\
\scriptsize (a) \hspace{3.7 cm}(b)\hspace{6 cm}
\end{center}
\caption{ Anomalous exponents vs hopping parameter $w$ and interaction strength $v_0$ for a step ladder ($v_F=1$) with the points on different poles and same side of the origin (a) $C_1$ (b) $D_1$.  The other exponents are independent of the hopping parameter.}
\label{SSDP}
\end{figure}

\begin{table}[h!]
\caption{Luttinger exponents $ U_{\nu_1,\nu_2}(\nu,\nu^{'};a;j) $ for $x_1$ and $x_2$ on opposite sides of the origin and on the same pole. Explicit expressions for the entries in the table are given in \hyperref[Appendix A]{Appendix A}.\\}
{ \begin{tabular}{c  c  c  c  c}
 \hline
   $ U_{\nu_1,\nu_1}(\nu,\nu^{'};a;j) $
   \mbox{  }& $\substack{ \nu = 1 ;\\\nu^{'} = 1 } $ & $\substack{\nu = -1 ;\\\nu^{'} = -1 }$ &
   $ \substack{\nu = 1 ; \\\nu^{'} = -1 } $ & $\substack{ \nu = -1 ; \\\nu^{'} = 1 } $  \\
   \hline
 $ \substack{\nu_1 = 1,\nu_2 = 1 ;\\a = h, j = 1  }$ &$A_2$ \hspace{0.6 cm}&\hspace{0.6 cm} $B_2$ \hspace{0.6 cm}&\hspace{0.6 cm} $C_2$\hspace{0.6 cm} &\hspace{0.6 cm} $D_2$\hspace{0.6 cm}  \\[4pt]
   $\substack{ \nu_1 = 1,\nu_2 = 1 ;\\ a = n, j = 1 } $ &$ 0.5 \mbox{  }  $ &0& $ -0.5 \mbox{  } $   & $ 0 \mbox{  }  $  \\[4pt]
 $\substack{ \nu_1 = 1,\nu_2 = 1 ; \\a = h, j = 2 } $  & $A_2$ & $B_2$ & $D_2$ & $C_2$ \\[4pt]
  $\substack{\nu_1 = 1,\nu_2 = 1 ;\\ a = n, j = 2 }$  &$ 0.5 \mbox{  }  $& 0&0&$ -0.5 \mbox{  } $\\[4pt]
  $\substack{ \nu_1 = -1,\nu_2 = -1 ; \\a = h, j = 1}  $  & $B_2$ & $A_2$ & $C_2$ & $D_2$  \\[4pt]
   $\substack{ \nu_1 = -1,\nu_2 = -1 ; \\a = n, j = 1 } $ & 0 &$ 0.5 \mbox{  }  $& $ -0.5 \mbox{  }  $ & $ 0 \mbox{  }  $  \\[4pt]
  $\substack{ \nu_1 = -1,\nu_2 =- 1 ;\\ a = h, j = 2}  $  & $B_2$ & $A_2$ & $D_2$ & $C_2$\\[4pt]
  $\substack{ \nu_1 = -1,\nu_2 = -1 ;\\ a = n, j = 2 } $&0&$ 0.5 \mbox{  }  $&0 & $-0.5$ \\[4pt]
  \hline
\end{tabular}}
\label{OSSP33}
\end{table}

\begin{table}[h!]
\caption{Luttinger exponents $ W_{\nu_1,\nu_2}(\nu,\nu^{'};a;j) $ for $x_1$ and $x_2$ on opposite sides of the origin and on different poles. Explicit expressions for the entries in the table are given in section \hyperref[Appendix A]{Appendix A}.\\}
{ \begin{tabular}{c  c  c  c  c}
 \hline
   $ W_{\nu_1,\nu_1}(\nu,\nu^{'};a;j) $
   \mbox{  }& $\substack{ \nu = 1 ;\\\nu^{'} = 1 } $ & $\substack{\nu = -1 ;\\\nu^{'} = -1 }$ &
   $ \substack{\nu = 1 ; \\\nu^{'} = -1 } $ & $\substack{ \nu = -1 ; \\\nu^{'} = 1 } $  \\
   \hline
 $ \substack{\nu_1 = 1,\nu_2 = 1 ;\\a = h, j = 1  }$ &$A_3$ \hspace{0.6 cm}&\hspace{0.6 cm} $B_3$ \hspace{0.6 cm}&\hspace{0.6 cm} $C_3$\hspace{0.6 cm} &\hspace{0.6 cm} $D_3$\hspace{0.6 cm}  \\[4pt]
   $\substack{ \nu_1 = 1,\nu_2 = 1 ;\\ a = n, j = 1 } $ &$ 0.5 \mbox{  }  $ &0& $ -0.5 \mbox{  } $   & $ 0 \mbox{  }  $  \\[4pt]
 $\substack{ \nu_1 = 1,\nu_2 = 1 ; \\a = h, j = 2 } $  & $A_3$ & $B_3$ & $D_3$ & $C_3$ \\[4pt]
  $\substack{\nu_1 = 1,\nu_2 = 1 ;\\ a = n, j = 2 }$  &$ 0.5 \mbox{  }  $& 0&0&$ -0.5 \mbox{  } $\\[4pt]
  $\substack{ \nu_1 = -1,\nu_2 = -1 ; \\a = h, j = 1}  $  & $B_3$ & $A_3$ & $C_3$ & $D_3$  \\[4pt]
   $\substack{ \nu_1 = -1,\nu_2 = -1 ; \\a = n, j = 1 } $ & 0 &$ 0.5 \mbox{  }  $& $ -0.5 \mbox{  }  $ & $ 0 \mbox{  }  $  \\[4pt]
  $\substack{ \nu_1 = -1,\nu_2 =- 1 ;\\ a = h, j = 2}  $  & $B_3$ & $A_3$ & $D_3$ & $C_3$\\[4pt]
  $\substack{ \nu_1 = -1,\nu_2 = -1 ;\\ a = n, j = 2 } $&0&$ 0.5 \mbox{  }  $&0 & $-0.5$ \\[4pt]
  \hline
\end{tabular}}
\label{OSDP44}
\end{table}
\begin{figure}[t!]
\begin{center}
\includegraphics[scale=0.325]{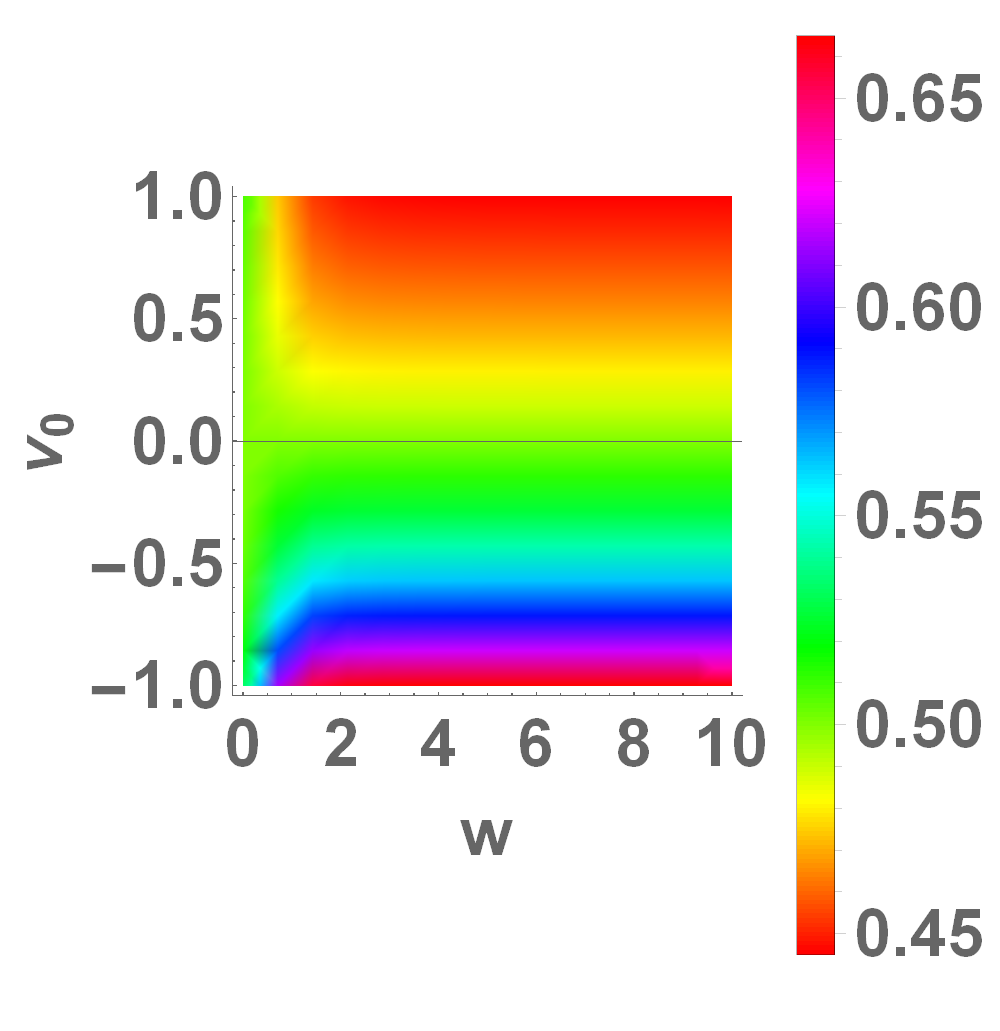}\hspace{0.1 cm}
\includegraphics[scale=0.335]{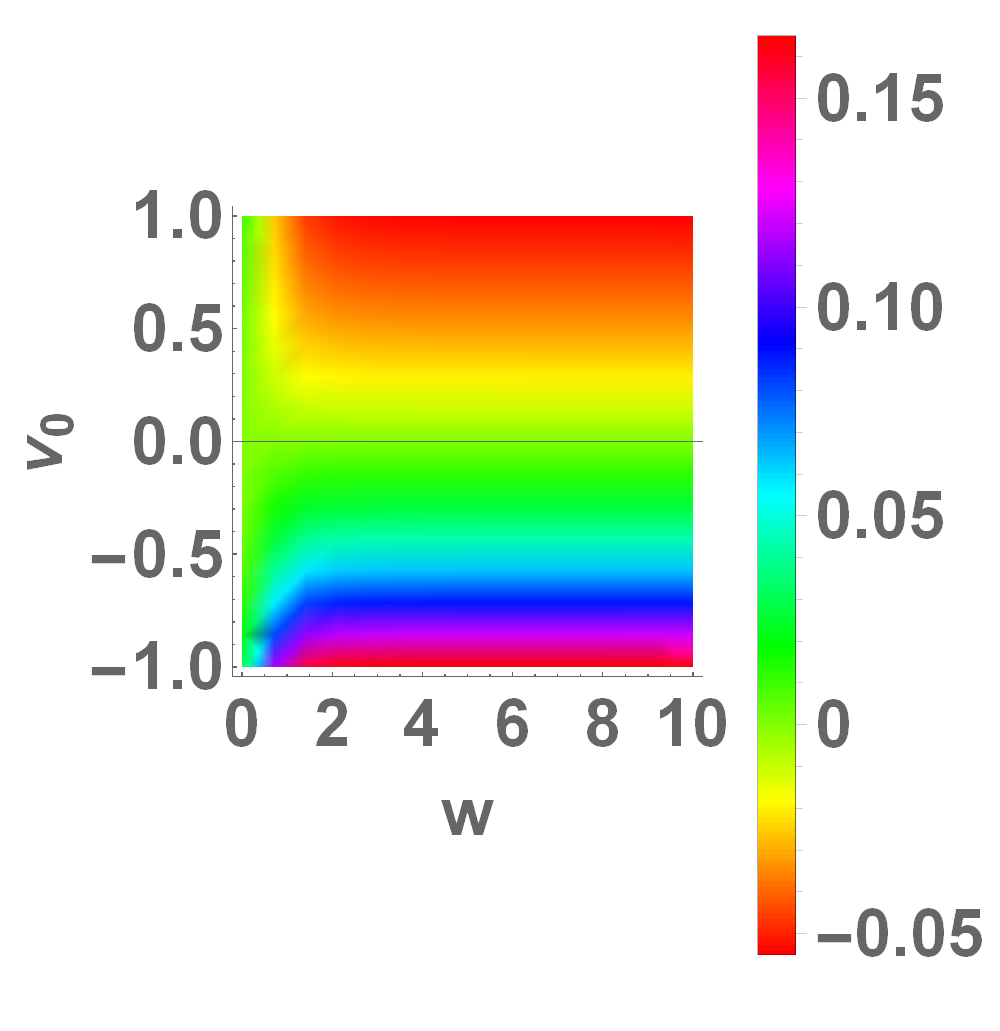}\hspace{0.1 cm}\\
\scriptsize (a) \hspace{3.6 cm}(b)\hspace{4.5 cm}
\end{center}
\caption{ Anomalous exponents vs hopping parameter $w$ and interaction strength $v_0$ for a step ladder ($v_F=1$) with the points on the same pole and different sides of the origin  (a) $A_2$ (b) $B_2$.  The other exponents are independent of the hopping parameter.}
\label{OSSP}
\end{figure}
\begin{figure}[h!]
\begin{center}
\includegraphics[scale=0.315]{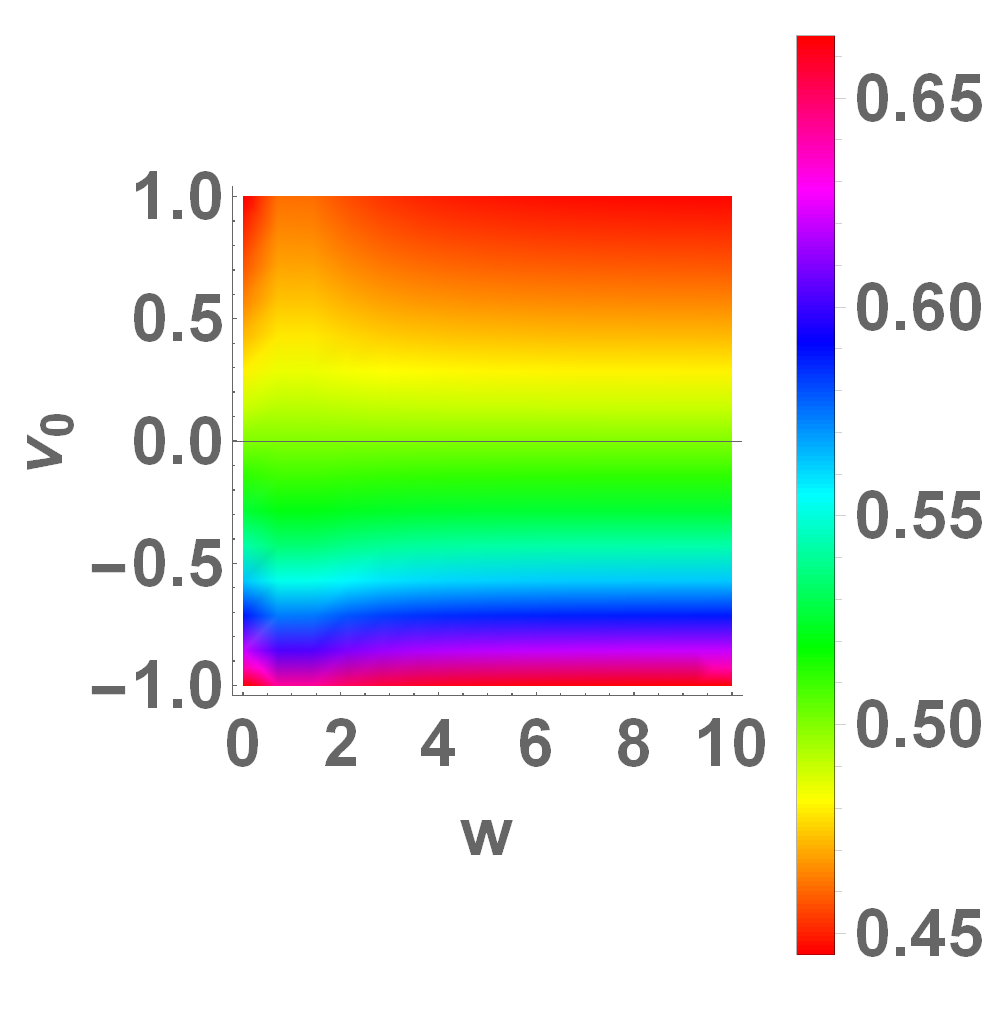}\hspace{0.1 cm}
\includegraphics[scale=0.325]{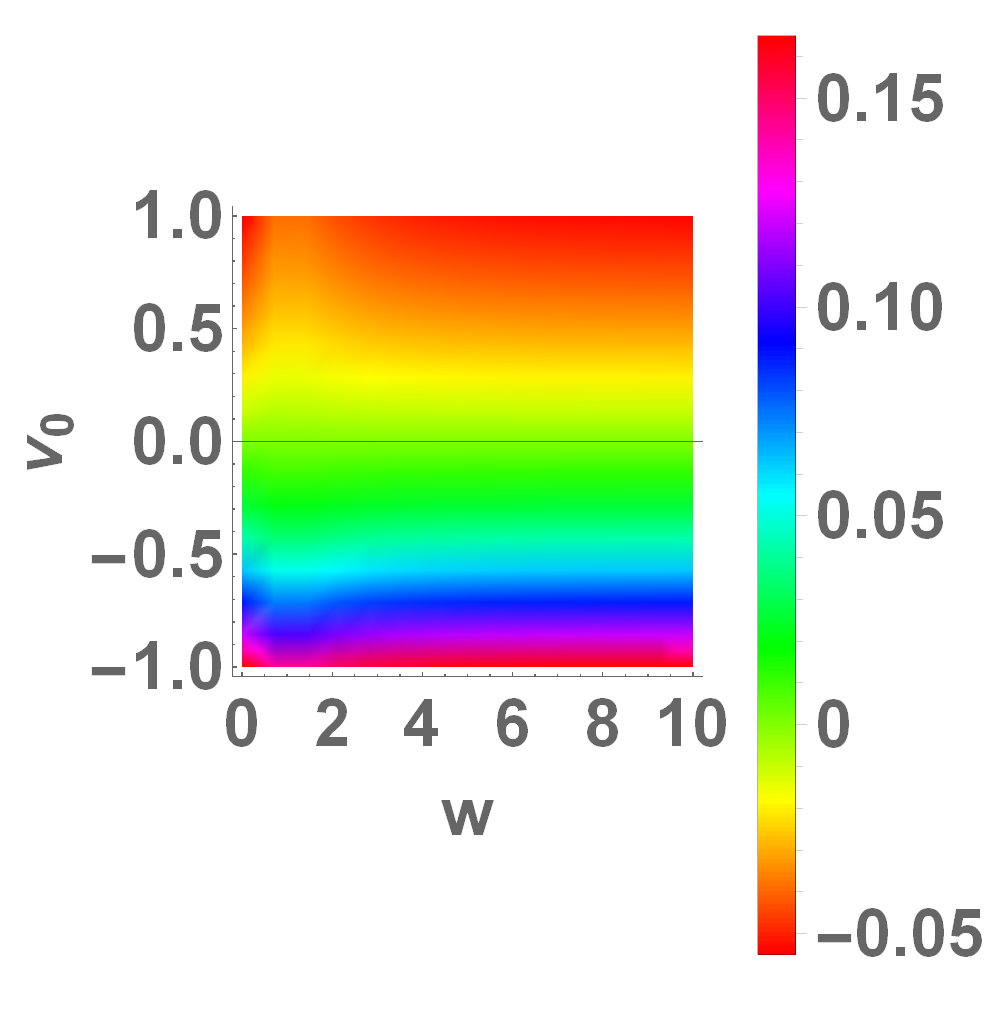}\hspace{0.1 cm}\\
\scriptsize (a) \hspace{3.6 cm}(b)\hspace{5 cm}
\end{center}
\caption{ Anomalous exponents vs hopping parameter $w$ and interaction strength $v_0$ for a step ladder ($v_F=1$) with the points on  opposite poles and different sides of the origin (a) $A_3$ (b) $B_3$.  The other exponents are independent of the hopping parameter.}
\label{OSDP}
\end{figure}
\normalsize
\subsubsection*{Anomalous exponents}
\label{LuttingerExpo}
The explicit expressions of the anomalous exponents mentioned in Table \ref{SSSP},  Table \ref{SSDP11},  Table \ref{OSSP33},  and Table \ref{OSDP44} are listed in \hyperref[Appendix A]{Appendix A}.
Some of the anomalous exponents are plotted as a function of hopping parameter and interaction parameter (taking empirical value of $v_F$ to be 1) in the Figures \ref{XYZ}, \ref{SSDP}, \ref{OSSP} and \ref{OSDP}. Only those exponents which have a dependence on both the hopping parameter and mutual interaction strength are plotted. Thus the exponents for spinons are omitted as they take only trivial values even in presence of interactions as it is the total density which couples to the short range mutual interactions. The key observations from the plots may be summarized as follows. When  both interactions and hopping are absent, all the exponents take on trivial values of zero or half (the other half comes from spinons to make the exponent unity). When hopping is absent but interactions are present, the exponents take the exact values as of the standard (homogeneous) Luttinger liquid \ref{ab} (limiting cases).

\section{Conductance}
\begin{figure}[h!]
\begin{center}
\includegraphics[scale=0.325]{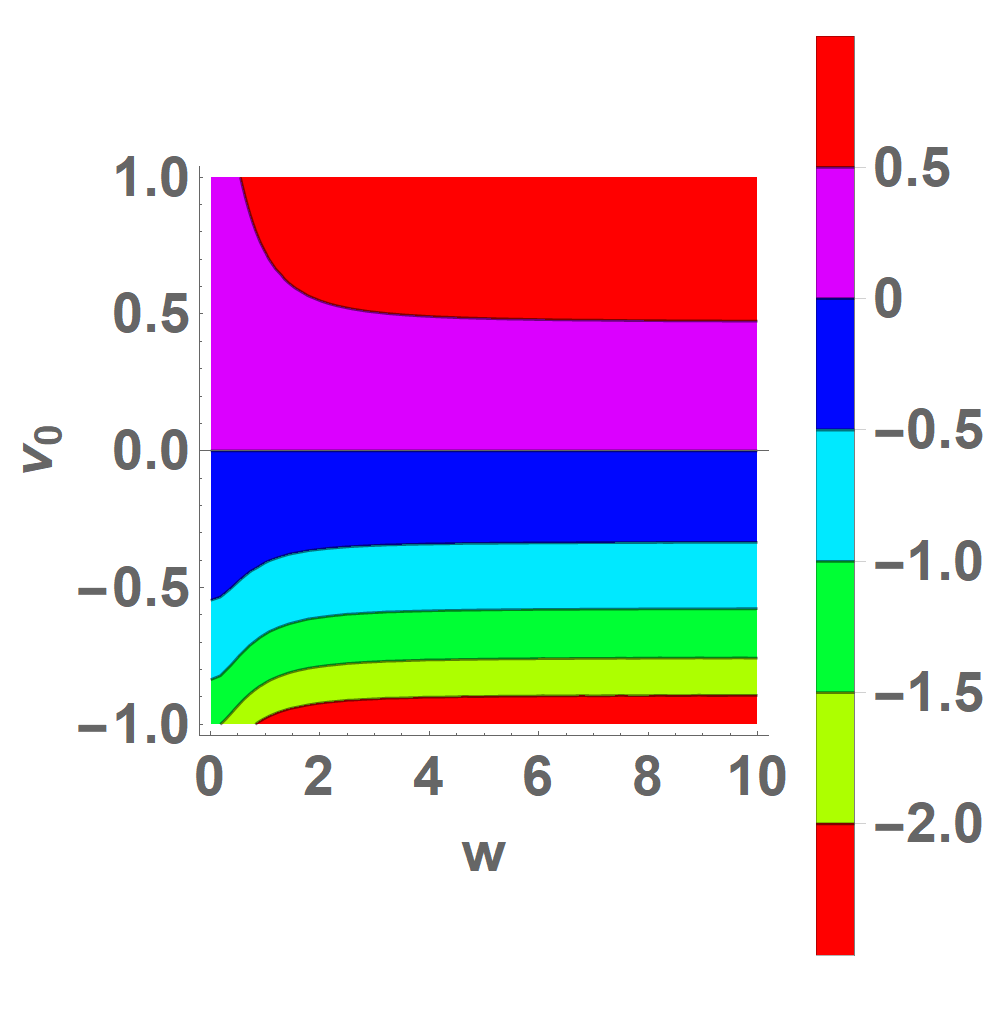}\hspace{0.1 cm}
\includegraphics[scale=0.325]{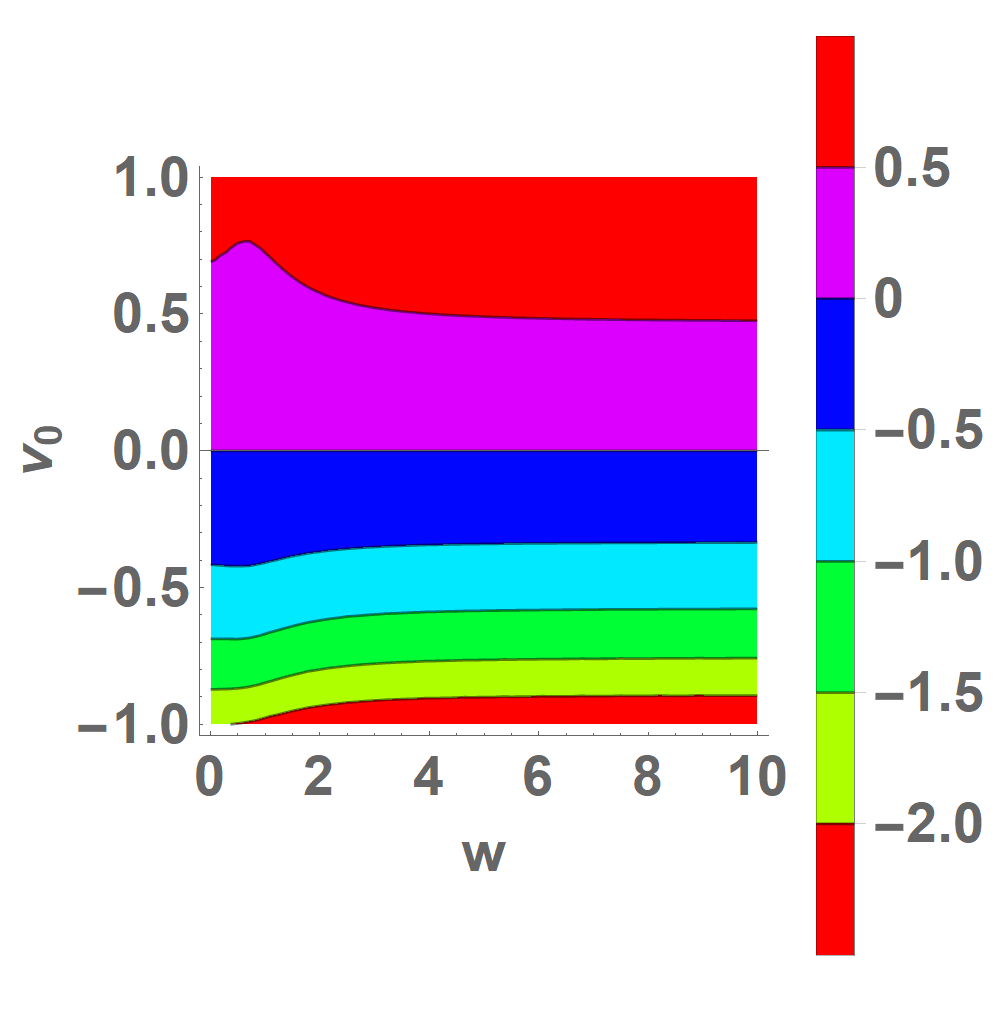}\hspace{0.2 cm}
\\
\scriptsize(a)\hspace{4cm}(b)
\end{center}
\caption{Contour plot showing the variation of the conductance exponent as a function of the hopping parameter `w' and the strength of mutual interaction `$v_0$' ($v_F=1$).  (a) For $\eta_1$ .(b) For $\eta_2 (=\eta_3)$.}
\label{eta}
\end{figure}

 Conductance may be thought of as the outcome of a tunneling experiment \cite{kane1992transport}. In this case the results depend on the length of the wire $ L $ and a cutoff $ L_{\omega} = \frac{ v_F }{ k_B T } $ that may be regarded either as inverse temperature or inverse frequency (former in case of d.c. conductance at finite temperature and latter in case of a.c. conductance at zero temperature). The result derived is the following:\\
\begin{equation}
G \sim \left( \frac{ L }{ L_{ \omega } }\right)^{\eta }  \mbox{  }
\label{GGEN}
\end{equation}
\begin{figure}[b!]
\begin{center}
\includegraphics[scale=0.4]{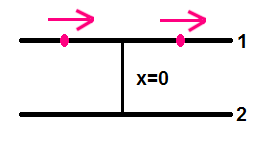}\hspace{0.2 cm}
\includegraphics[scale=0.4]{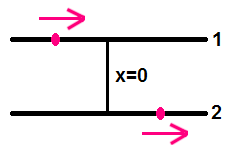}\hspace{0.2 cm}
\includegraphics[scale=0.4]{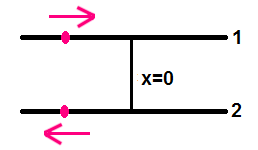}\hspace{0.2 cm}\\
\scriptsize(a)\hspace{2.5 cm}(b)\hspace{2.5 cm}(c)\\
\end{center}
\caption{Potential difference applied across different ends to measure the conductance.}
\label{Conductance}
\end{figure}
\\ \mbox{  } \\
(a) When the voltage difference is applied to the ends of same Luttinger Liquid, then
\[
\eta_1=2(4+\gamma_1+\gamma_2-A_2-B_2-2D_2-\frac{1}{2})
\]
\\ \mbox{  } \\
(b) When the voltage difference is applied to the ends of different Luttinger Liquids on opposite sides of the origin, then
\[
\eta_2=2(4+\gamma_1+\gamma_2-A_3-B_3-2D_3- \frac{1}{2})
\]
\\ \mbox{  } \\
(c) When the voltage difference is applied to the ends of different Luttinger Liquids on the same side of the origin, then
\[
\eta_3=2(4+\gamma_1+\gamma_2-2B_1-C_1-D_1- \frac{1}{2})
\]
\\ \mbox{  } \\
 The explicit values of A's, B's, etc. are given in \hyperref[Appendix A]{Appendix A} and the derivation of the exponents is provided in \hyperref[Appendix C]{Appendix C}. It is to be noted that $\eta_2=\eta_3$ (see Fig. \ref{equivalence} for a diagrammatic explanation). Fig. \ref{eta} shows the variation of the conductance exponent as a function of the hopping parameter and the strength of mutual interaction. It is seen that for attractive interactions ($v_0<0$), the conductance exponents are negative ($\eta <0$) indicating that the conductance diverges at low temperature as a power law. On the other hand, for repulsive interactions ($v_0>0$), the conductance exponents are positive ($\eta >0$ ) and thus conductance vanishes at low temperature. When mutual interactions are absent  ($v_0=0$), then the conductance exponent vanishes ($\eta=0$) which indicates that in such cases, the conductance is independent of temperature.
\section{Friedel Oscillations}
The presence of a localized impurity that weakly couples to the fermions causes Friedel oscillations, which are  the rapid spatial variation ($ \sim  e^{   2i k_F x }$) of the otherwise homogeneous density profile in a Luttinger liquid. In the Kubo formalism it is given as the density-density correlation function \cite{friedel1958metallic, tutto1988theory}. Egger and Grabert have studied Friedel oscillations in a Luttinger liquid with arbitrary interactions and arbitrary strengths of impurities \cite{egger1995friedel}.  Consider the expression in equation (\ref{INPUT1}). Define $ {\tilde{\rho}}_f \equiv \rho_f -<\rho_f> $.
Using the non standard harmonic analysis for the one step ladder, we write the fast part of the density operator as follows:\\
\begin{equation}
\begin{aligned}
\rho_f^{i}&(x,\sigma,t) \equiv \psi^{\dagger}_{L,i}(x,\sigma,t) \psi_{R,i}(x,\sigma,t)
\\
 = &\sum_{\lambda = 0,1 } \sum_{ \gamma = \pm 1}
  \theta(\gamma x) \mbox{  }R^{i}_{ \lambda,\gamma }(\sigma)\mbox{       } e^{ 2 \pi i \int^{x}_{sgn(x)\infty } dy \mbox{      }\rho^{i}_{s}(y,\sigma,t) }\mbox{      }\\
&\mbox{      } \mbox{      }   e^{ 2 \pi i \lambda\int^{x}_{sgn(x)\infty } dy \mbox{      }  ( \rho^{ {\bar{i}} }_{s}(y,\sigma,t)  +
  \rho^{ 1 }_{s}(-y,\sigma,t) +\rho^{ 2 }_{s}(-y,\sigma,t) )  }
\end{aligned}
\end{equation}

The prescription for choosing $ \lambda_i $ in equation (\ref{PSII}) leads  to the unambiguous conclusion that,\footnotesize
\begin{equation}
\begin{aligned}
&\Big\langle T\mbox{  } {\tilde{\rho}}^i_f(X_1)  {\tilde{\rho}}^j_f(X_2)\Big\rangle \sim \\
&\mbox{  }  ( Exp[ \sum_{ \substack{\nu,\nu^{'} = \pm 1 \\ a = h,n }   }\Gamma(\nu,\nu^{'};a) \mbox{  } Log[ (\nu x_1 - \nu^{'} x_2 ) - v_a (t_1-t_2) ] ]-1 ) \\
&\Big\langle T\mbox{  } {\tilde{\rho}}^i_f(X_1)  {\tilde{\rho}}^{j*}_f(X_2)\Big\rangle \sim \\
&\mbox{  }( Exp[ - \sum_{ \substack{\nu,\nu^{'} = \pm 1 \\ a = h,n }   }\Gamma(\nu,\nu^{'};a) \mbox{  } Log[ (\nu x_1 - \nu^{'} x_2 ) - v_a (t_1-t_2) ] ]-1 ) \\
\end{aligned}
\end{equation}
\normalsize One should remember that this (use of tilde ``$ \sim $") really means that the time derivative of the logarithms of both sides are equal to each other. The values of the anomalous scaling exponents $\Gamma(\nu,\nu^{'};a)$ can be obtained from the expression below.\\
\begin{equation}
\begin{aligned}\footnotesize
\Gamma(\nu,\nu^{'};a) =& \left(\frac{v_F}{2v_h}\mbox{ }\delta_{a,h}+\frac{1}{2} \mbox{ }\delta_{a,n} \right)\mbox{ } (\delta_{\nu,\nu'}-\delta_{\nu,-\nu'})
\end{aligned}
\end{equation}

\section{Limiting checks}
\label{ab}
\subsection{Non-interacting case}
The obvious limiting check is to switch off the inter-particle interactions ($v_0=0$) and then compare with the respective single particle Green  functions obtained using Fermi algebra. In such a case, the holon velocity is equal to the Fermi velocity ($v_h \to v_F$) and all the subcases of the interacting two point functions in equations (\ref{SS1}, \ref{SS2}, \ref{OS1}, \ref{OS2})  will become identical to equation (\ref{INPUT1}). The details of this calculation can be found in \hyperref[Appendix B]{Appendix B}.
Similarly, switching off interactions in equations (\ref{DDOSI1}) and (\ref{DDOSI2}) leads to the the non interacting density density correlation functions as given by equation (\ref{INPUT3}).
\subsection{No hopping}
\noindent  When the hopping parameter vanishes ($w=0$) then from equation (\ref{SS1}), it is obvious that  \\
\small
\[
 <T \mbox{  } \psi_R(x_1,t_1)\psi^{\dagger}_L(x_2,t_2)> = <T \mbox{  } \psi_L(x_1,t_1)\psi^{\dagger}_R(x_2,t_2)> =  0
\]\normalsize
 Moreover in this situation,
\footnotesize
\begin{equation*}
\begin{aligned}
&P=\frac{(v_h+v_F)^2}{8 v_h v_F} \mbox{ }; \hspace{.5 cm}Q=\frac{(v_h-v_F)^2}{8 v_h v_F} \mbox{ }; \hspace{.5 cm}X= 0  \mbox{ };
\hspace{.5 cm}\gamma_1=0
\end{aligned}
\end{equation*}
\normalsize
Hence the only non-vanishing parts of the NCBT two-point function for points on the same side the origin are,
\footnotesize
\begin{equation}
\begin{aligned}
\Big\langle T&\psi_R(x_1,\sigma_1,t_1)\psi_R^{\dagger}(x_2,\sigma_2,t_2)\Big\rangle=\frac{i}{2\pi}\mbox{  } e^{-\frac{1}{2} \log{[(x_1-x_2)-v_F(t_1-t_2)]}}\\
&\hspace{-0.7cm}e^{-\frac{(v_h+v_F)^2}{8 v_h v_F} \log{[(x_1-x_2)-v_h(t_1-t_2)]}}e^{-\frac{(v_h-v_F)^2}{8 v_h v_F}  \log{[-(x_1-x_2)-v_h(t_1-t_2)]}}\\
\Big\langle T&\psi_L(x_1,\sigma_1,t_1)\psi_L^{\dagger}(x_2,\sigma_2,t_2)\Big\rangle= \frac{i}{2\pi}\mbox{  }e^{-\frac{1}{2} \log{[-(x_1-x_2)-v_F(t_1-t_2)]}}\\
&\hspace{-0.7cm}e^{-\frac{(v_h-v_F)^2}{8 v_h v_F}  \log{[(x_1-x_2)-v_h(t_1-t_2)]}}e^{-\frac{(v_h+v_F)^2}{8 v_h v_F} \log{[-(x_1-x_2)-v_h(t_1-t_2)]}}
\label{Standard}
\end{aligned}
\end{equation}
\normalsize
These are precisely the standard Luttinger liquid two-point functions of a translationally invariant system.
 For points on the opposite sides of the origin, the most singular forms of the asymptotic Green functions have a discontinuous dependence on the hopping parameter $ w $ at $ w = 0 $ when mutual interactions between fermions are present.
\subsection{Mandatory hopping}
\begin{figure}[h!]
\begin{center}
\includegraphics[scale=0.42]{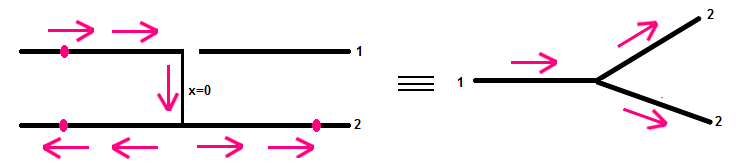}\hspace{0.2 cm}
\end{center}
\caption{An electron after hopping from pole 1 when just reaches pole 2 has equal probabilities going to either sides due to identical environments.}
\label{equivalence}
\end{figure}

The other extreme limit is to allow the hopping parameter tend to infinity ($w \to \infty$), so that the particle compulsorily travels through the connecting link without going to the other side of the same pole. In such case, the two point functions of the `case III: Same pole opposite sides'  (equation (\ref{OS1})) vanishes. Furthermore,
\begin{equation*}
\begin{aligned}
&A_1=C_3=-\frac{v_h+v_F}{4v_h};
&B_1=D_3=\frac{v_h-v_F}{4v_h}\\
&C_1=A_3=\frac{v_h+v_F}{4v_h};
&D_1=B_3=-\frac{v_h-v_F}{4v_h}\\
\end{aligned}
\end{equation*}
Therefore from equation (\ref{SS2}) and (\ref{OS2}) we have the following (see Fig. \ref{equivalence}):
\begin{equation*}
\begin{aligned}
\Big\langle T\mbox{  }&\psi^1_{R}(x_1,t_1,\sigma_1)\psi_{R}^{2\dagger}(x_2,t_2,\sigma_2)\Big\rangle_{\text{opposite sides}}\\
&=\Big\langle T\mbox{  }\psi^1_{R}(x_1,t_1,\sigma_1)\psi_{L}^{2\dagger}(-x_2,t_2,\sigma_2)\Big\rangle _{\text{same side}} \\
\end{aligned}
\end{equation*}
\subsection{Far away from hopping site}
Consider the situation when $ x_1 > 0,\mbox{ } x_2 > 0 $. Set $ R_{cm} = (x_1+x_2)/2 $ and $ y = x_1-x_2 $. This means
 $ x_1 = R_{cm} + \frac{y}{2} $ and $ x_2 = R_{cm} - \frac{y}{2} $. If $ y $ is held fixed and  $ R_{cm} \rightarrow \infty $, then this depicts the region far away from the hopping site. In such a situation it is expected that Green's functions such as $ <T \mbox{  }\psi_R(x_1,t_1) \psi^{\dagger}_R(x_2,t_2) > $ and
    $ <T \mbox{  }\psi_L(x_1,t_1) \psi^{\dagger}_L(x_2,t_2) > $ to be immune to the presence or absence of the hopping sites. However the parts
     $ <T \mbox{  }\psi_R(x_1,t_1) \psi^{\dagger}_L(x_2,t_2) > $ and
    $ <T \mbox{  }\psi_L(x_1,t_1) \psi^{\dagger}_R(x_2,t_2) > $ that are non-zero only because of the hopping site have no such restriction.
     In passing it is noted that while the opposite choice viz. holding $ R_{cm} $ fixed while making $ y \rightarrow \infty $ also makes the two points far from the impurity, since the region where the impurity is present has to be traversed, this Green function certainly will not be immune to the presence of the impurity.
\footnotesize
\begin{equation}
\begin{aligned}
\Big\langle T\psi_R(x_1&,\sigma_1,t_1)\psi_R^{\dagger}(x_2,\sigma_2,t_2)\Big\rangle\\
\sim &e^{\gamma_1 \log{[4R_{cm}^2-y^2]}}e^{-\frac{1}{2} \log{[y-v_F(t_1-t_2)]}}\\
&e^{-P \log{[y-v_h(t_1-t_2)]}}e^{-Q \log{[-y-v_h(t_1-t_2)]}}\\
&e^{-X \log{[2R_{cm}-v_h(t_1-t_2)]}} e^{-X \log{[-2R_{cm}-v_h(t_1-t_2)]}}\\
\Big\langle T\psi_L(x_1&,\sigma_1,t_1)\psi_L^{\dagger}(x_2,\sigma_2,t_2)\Big\rangle\\
\sim &e^{\gamma_1 \log{[4R_{cm}^2-y^2]}}e^{-\frac{1}{2} \log{[-y-v_F(t_1-t_2)]}}\\
&e^{-Q \log{[y-v_h(t-t_2)]}}e^{-P \log{[-y-v_h(t_1-t_2)]}}\\
&e^{-X \log{[2R_{cm}-v_h(t_1-t_2)]}} e^{-X \log{[-2R_{cm}-v_h(t_1-t_2)]}}
\label{samesidefull3}
\end{aligned}
\end{equation}
\normalsize
where,
\small
\begin{equation*}
\begin{aligned}
P=&\frac{(v_F+v_h)^2}{8 v_F v_h};\hspace{0.5cm}Q=\frac{(v_F-v_h)^2}{8 v_F v_h};\hspace{0.5cm}\gamma_1=X;\\\hspace{0.5cm}
X=&-\frac{(v_F-v_h)(v_F+v_h)w^2(v_F^4-v_F^2v_h^2+v_Fv_hw^2+w^4)}{8 v_Fv_h(v_Fv_h+w^2)(v_F^4+2v_Fv_hw^2+w^4)}\\
\end{aligned}
\end{equation*}
\normalsize
Now we have $ P - Q = \frac{1}{2} $ and $ \gamma_1  =X $. Performing said limit $ R_{cm} \rightarrow \infty $ and holding everything else fixed reduces Eq.(\ref{samesidefull3}) to,
\begin{equation}
\begin{aligned}
\Big\langle T\psi_R(x_1,&\sigma_1,t_1)\psi_R^{\dagger}(x_2,\sigma_2,t_2)\Big\rangle\\
\sim &e^{-\frac{1}{2} \log{[y-v_F(t_1-t_2)]}}
e^{-\frac{1}{2} \log{[y-v_h(t_1-t_2)]}}\hspace{0cm}\\
&e^{-Q \log{[y-v_h(t_1-t_2)]}}e^{-Q \log{[-y-v_h(t_1-t_2)]}}\\
\Big\langle T\psi_L(x_1,&\sigma_1,t_1)\psi_L^{\dagger}(x_2,\sigma_2,t_2)\Big\rangle \\
\sim& e^{-\frac{1}{2} \log{[-y-v_F(t_1-t_2)]}}e^{-\frac{1}{2} \log{[-y-v_h(t_1-t_2)]}}\\
&e^{-Q \log{[-y-v_h(t_1-t_2)]}}e^{-Q \log{[y-v_h(t_1-t_2)]}}
\label{GREENFAR}
\end{aligned}
\end{equation}
Since $ Q=\frac{(v_h-v_F)^2}{8 v_h v_F}  $, Eq.(\ref{GREENFAR}) is precisely the Green function of a homogeneous Luttinger liquid.
In other words these Green's functions are immune to the presence of the impurity.
\\ \mbox{   } \\

\section{Conclusions}
The ``non-chiral" bosonization technique employed in this work is based on a non-standard harmonic analysis of the rapidly varying parts of the density fields appropriate for the study of strongly inhomogeneous systems such as the one-step ladder. It is used to obtain closed analytical formulas for the two-point function and four point functions relevant to Friedel oscillations. Unlike g-ology based methods, the present approach treats the source of inhomogeneities exactly. The analytical expressions for the correlation functions written down are nothing but the resummation of the most singular parts of the RPA terms in an expansion in powers of the mutual interaction using Fermi algebra.  To further validate the results, various limiting cases are cross checked. Finally the conductance of the system is calculated for the biasing voltage applied across different ends of the ladder.

\section*{Funding}
A part of this work was done with financial support from Department of Science and Technology, Govt. of India
DST/SERC: SR/S2/CMP/46 2009.


\section*{APPENDICES}

\setcounter{section}{1}
\setcounter{equation}{0}
\renewcommand{\theequation}{\Alph{section}.\arabic{equation}}
\section*{APPENDIX A:  Anomalous Exponents}
\label{Appendix A}

The explicit expressions of the anomalous exponents mentioned in Table \ref{SSSP},  Table \ref{SSDP11},  Table \ref{OSSP33},  and Table \ref{OSDP44} are listed below. For compactness, all the other exponents are written in terms of the following five exponents.
\footnotesize
\begin{equation}
\begin{aligned}
Q=&\frac{(v_F-v_h)^2}{8 v_F v_h}\mbox{  };\mbox{ }B_1=\frac{v_h-v_F}{4v_h}\mbox{ }\\
X=&-\frac{(v_F-v_h)(v_F+v_h)w^2(v_F^4-v_F^2v_h^2+v_Fv_hw^2+w^4)}{8 v_Fv_h(v_Fv_h+w^2)(v_F^4+2v_Fv_hw^2+w^4)}\\
C_1=&\frac{(v_h+v_F)(2v_F^4+v_h(3v_F+v_h)w^2+2w^4)}{8v_h(v_F^4+2v_Fv_hw^2+w^4)}\mbox{  };\mbox{ }\\
A_2=&-\frac{(v_h+v_F)}{8}\bigg(-\frac{2}{v_h}-\frac{(v_h-v_F)}{v_Fv_h+w^2}+\frac{w^2(v_h-v_F)}{v_F^4+2v_Fv_hw^2+w^4} \bigg)\\
\end{aligned}
\end{equation}

\noindent Note that, when\mbox{  } $v_h=v_F$\mbox{  },\mbox{  } $Q=X=B_1=0$\mbox{  };\mbox{ }$C_1=A_2=\frac{1}{2}$.\\
Also, when \mbox{  }$w=0$\mbox{  },\mbox{  }$Q=\frac{(v_F-v_h)^2}{8 v_F v_h}$\mbox{  };\mbox{ }$B_1=\frac{v_h-v_F}{4v_h}$\mbox{  };\mbox{ }$X=0$\mbox{  };\mbox{ }$\\
C_1=\frac{v_h+v_F}{4v_h}$\mbox{  };\mbox{ }$A_2=\frac{(v_h+v_F)^2}{8v_Fv_h}$\mbox{  }\\

\begin{bf} \noindent Homogeneous Exponents\end{bf} 
\begin{equation}
\begin{aligned}
&\gamma_1=X\mbox{  };\mbox{ }
&\gamma_2=X+6B_1-3\\
\end{aligned}
\end{equation}When $v_h=v_F$, $\gamma_1=0$ and $\gamma_2=-3$.\\ \mbox{  } \\
\begin{bf} \noindent Case I : $x_1$ and $x_2$ on the same side of the origin and same pole\end{bf} \\
\begin{equation}
\begin{aligned}
P=  Q+\frac{1}{2} \mbox{  };\mbox{ }
S= \frac{Q}{B_1}\left(\frac{1}{2}-B_1\right) ;\mbox{ }
Y= X-B_1+\frac{1}{2} \mbox{  };\mbox{ }
Z=  X-B_1\\
\end{aligned}
\end{equation}
\\ \mbox{  } \\
 \begin{bf}Case II : $x_1$ and $x_2$ on same side of the origin and different pole\end{bf}
\begin{equation}
\begin{aligned}
A_1= B_1-\frac{1}{2}  \mbox{  };\mbox{ }
D_1= C_1-\frac{1}{2} \mbox{  };\mbox{ }
\end{aligned}
\end{equation}
\\ \mbox{  } \\
 \begin{bf}Case III : $x_1$ and $x_2$ on the opposite sides of the origin and on the same pole\end{bf} \\
\begin{equation}
\begin{aligned}
B_2= A_2-\frac{1}{2}  \mbox{  };\mbox{ }
C_2= B_1-\frac{1}{2} \mbox{  };\mbox{ }
D_2=B_1  \mbox{  }
\end{aligned}
\end{equation}
\\ \mbox{  } \\
 \begin{bf}Case IV : $x_1$ and $x_2$ on the opposite sides of the origin and on different poles\end{bf} \\
\begin{equation}
\begin{aligned}
A_3=C_1 \mbox{  };\mbox{ }
B_3=C_1-\frac{1}{2} \mbox{  };\mbox{ }
C_3= B_1-\frac{1}{2} \mbox{  };\mbox{ }
D_3= B_1 \mbox{  };\mbox{ }
\end{aligned}
\end{equation}
\setcounter{section}{2}
\setcounter{equation}{0}
\renewcommand{\theequation}{\Alph{section}.\arabic{equation}}
\section*{APPENDIX B:  Limiting case checks}
\label{Appendix B}
\footnotesize
\subsection{ Non interacting case}
\begin{bf} \noindent Homogeneous Exponents \end{bf} \\
\begin{equation}
\begin{aligned}
&\gamma_1:=0;&\gamma_2:=-3\\
\end{aligned}
\end{equation}

\begin{bf} \noindent Case I : $x_1$ and $x_2$ on the same side of the origin and same pole\end{bf} \\
\begin{equation}
\begin{aligned}
P=&\frac{1}{2}\mbox{ };\mbox{ }     Q=0\mbox{  };\mbox{ }      S=0 \mbox{  };\mbox{ }   X=0\mbox{  };\mbox{ }    Z=0\mbox{  };\mbox{ }
Y=\frac{1}{2}\\
\end{aligned}
\end{equation}


 \noindent\begin{bf}Case II : $x_1$ and $x_2$ on same side of the origin and different pole\end{bf}

\begin{equation}
\begin{aligned}
A_1=&-\frac{1}{2}\mbox{ };\mbox{ }     B_1=0\mbox{  };\mbox{ }      D_1=0 \mbox{  };\mbox{ }
C_1=\frac{1}{2}\mbox{  }\\
\end{aligned}
\end{equation}

\noindent\begin{bf}Case III : $x_1$ and $x_2$ on the opposite sides of the origin and on the same pole\end{bf} \\

\begin{equation}
\begin{aligned}
C_2=&-\frac{1}{2}\mbox{ };\mbox{ }     D_2=0\mbox{  };\mbox{ }      B_2=0 \mbox{  }\mbox{ }
A_2=\frac{1}{2}\mbox{  }\\
\end{aligned}
\end{equation}

\noindent\begin{bf}Case IV : $x_1$ and $x_2$ on the opposite sides of the origin and on different poles\end{bf} \\

\begin{equation}
\begin{aligned}
C_3=&-\frac{1}{2}\mbox{ };\mbox{ }     D_3=0\mbox{  };\mbox{ }      B_3=0 \mbox{  };\mbox{ }
A_3=\frac{1}{2}\mbox{  }\\
\end{aligned}
\end{equation}
\normalsize

Using the above exponents  in all the subcases of the interacting two point functions in equations (\ref{SS1}, \ref{SS2}, \ref{OS1}, \ref{OS2}), one recovers the non-interacting Green functions  as given by equation (\ref{INPUT1}). For example, one of the subcases (same side and same pole as in equation (\ref{SS1})) is explicitly shown.

\footnotesize

\begin{equation*}
\begin{aligned}
\Big\langle T\psi_R&(x_1,\sigma_1,t_1)\psi_R^{\dagger}(x_2,\sigma_2,t_2)\Big\rangle\\
=&\frac{i}{2\pi}\mbox{  }e^{\gamma_1\log{[4x_1x_2]|}} e^{-\frac{1}{2} \log{[(x_1-x_2)-v_F(t_1-t_2)]}}\\
&e^{-P \log{[(x_1-x_2)-v_h(t_1-t_2)]}}e^{-Q \log{[-(x_1-x_2)-v_h(t_1-t_2)]}}\\
&e^{-X \log{[(x_1+x_2)-v_h(t_1-t_2)]}}e^{-X \log{[-(x_1+x_2)-v_h(t_1-t_2)]}}\\
=&\frac{i}{2\pi}\mbox{  } e^{-\log{[(x_1-x_2)-v_F(t_1-t_2)]}}\\
=&\frac{i}{2\pi}\mbox{  }\frac{1}{(x_1-x_2)-v_F(t_1-t_2)}\\
\end{aligned}
\end{equation*}
\begin{equation*}
\begin{aligned}
\Big\langle T\psi_L&(x_1,\sigma_1,t_1)\psi_L^{\dagger}(x_2,\sigma_2,t_2)\Big\rangle\\
=&\frac{i}{2\pi}\mbox{  }e^{\gamma_1\log{[4x_1x_2]|}} e^{-\frac{1}{2} \log{[-(x_1-x_2)-v_F(t_1-t_2)]}}\\
&e^{-Q \log{[(x_1-x_2)-v_h(t_1-t_2)]}}e^{-P \log{[-(x_1-x_2)-v_h(t_1-t_2)]}}\\
&e^{-X \log{[(x_1+x_2)-v_h(t_1-t_2)]}}e^{-X \log{[-(x_1+x_2)-v_h(t_1-t_2)]}}\\
=&\frac{i}{2\pi}\mbox{  } e^{-\log{[-(x_1-x_2)-v_F(t_1-t_2)]}}\\
=&\frac{i}{2\pi}\mbox{  }\frac{1}{-(x_1-x_2)-v_F(t_1-t_2)}\\
\end{aligned}
\end{equation*}
\begin{equation*}
\begin{aligned}
\Big\langle T\psi_R&(x_1,\sigma_1,t_1)\psi_L^{\dagger}(x_2,\sigma_2,t_2)\Big\rangle\\
=\frac{i}{2\pi}& \frac{w^2}{w^2+v_F^2}
\frac{e^{\gamma_1\log{[2x_1]|}}e^{(3+\gamma_2)\log{[2x_2]|}} +e^{(1+\gamma_3)\log{[2x_1]|}}e^{\gamma_1\log{[2x_2]|}}}{2}\\
&e^{- \frac{1}{2} \log{[(x_1+x_2)-v_F(t_1-t_2)]}}\\
&e^{-S \log{[(x_1-x_2)-v_h(t_1-t_2)]}}e^{-S \log{[-(x_1-x_2)-v_h(t_1-t_2)]}}\\
&e^{-Y \log{[(x_1+x_2)-v_h(t_1-t_2)]}}e^{-Z \log{[-(x_1+x_2)-v_h(t_1-t_2)]}}\\
=&\frac{i}{2\pi}\frac{w^2}{w^2+v_F^2} \mbox{  } e^{-\log{[(x_1+x_2)-v_F(t_1-t_2)]}}\\
=&\frac{i}{2\pi}\frac{w^2}{w^2+v_F^2} \mbox{  }\frac{1}{(x_1+x_2)-v_F(t_1-t_2)}\\
\end{aligned}
\end{equation*}
\begin{equation*}
\begin{aligned}
\Big\langle T\psi_L&(x_1,\sigma_1,t_1)\psi_R^{\dagger}(x_2,\sigma_2,t_2)\Big\rangle\\
=\frac{i}{2\pi}& \frac{w^2}{w^2+v_F^2}
\frac{e^{\gamma_1\log{[2x_1]|}}e^{(3+\gamma_2)\log{[2x_2]|}} +e^{(1+\gamma_3)\log{[2x_1]|}}e^{\gamma_1\log{[2x_2]|}}}{2}\\
&e^{- \frac{1}{2} \log{[-(x_1+x_2)-v_F(t_1-t_2)]}} \\
&e^{-S \log{[(x_1-x_2)-v_h(t_1-t_2)]}}e^{-S \log{[-(x_1-x_2)-v_h(t_1-t_2)]}}\\
&e^{-Z \log{[(x_1+x_2)-v_h(t_1-t_2)]}}e^{-Y \log{[-(x_1+x_2)-v_h(t_1-t_2)]}}\\
=&\frac{i}{2\pi}\frac{w^2}{w^2+v_F^2} \mbox{  } e^{-\log{[-(x_1+x_2)-v_F(t_1-t_2)]}}\\
=&\frac{i}{2\pi}\frac{w^2}{w^2+v_F^2} \mbox{  }\frac{1}{-(x_1+x_2)-v_F(t_1-t_2)}\\
\end{aligned}
\end{equation*}

\setcounter{section}{3}
\setcounter{subsection}{0}
\setcounter{equation}{0}
\renewcommand{\theequation}{\Alph{section}.\arabic{equation}}
\section*{APPENDIX C:  Conductance exponent derivation}
\label{Appendix C}

\subsection{Voltage applied across the same pole but opposite sides of the origin}
\normalsize
Consider the general Green function derived earlier for $ xx^{'} < 0 $ and same pole (Case III). From that it is possible to conclude
($ W = g^{1,1}_{1,-1}(1,1)\theta(x)\theta(-x')+g^{1,1}_{-1,1}(1,1)\theta(-x)\theta(x') $),
\footnotesize
\begin{equation*}
\begin{aligned}
<T\psi_R(&x_1,\sigma_1,t_1)\psi_R^{\dagger}(x_2,\sigma_2,t_2)>=\frac{e^{\gamma_1\log{[2x_1]}}e^{(3+\gamma_2)\log{[2x_2]}}}{2(x_1+x_2)}\mbox{  } \\
&\mbox{   }W\mbox{  }e^{-E \log{[(x_1-x_2)-v_F(t_1-t_2)]}}e^{\frac{1}{2} \log{[(x_1+x_2)-v_F(t_1-t_2)]}}\\
&e^{-A_2\log{[(x_1-x_2)-v_h(t_1-t_2)]}}e^{-B_2 \log{[(x_1-x_2)+v_h(t_1-t_2)]}}\\
&e^{-C_2\log{[(x_1+x_2)-v_h(t_1-t_2)]}} e^{-D_2 \log{[(x_1+x_2)+v_h(t_1-t_2)]}}\\
&+\frac{e^{(3+\gamma_2)\log{[2x_1]}}e^{\gamma_1\log{[2x_2]}} }{2(x_1+x_2)}\mbox{  } \\
&W\mbox{  }e^{-E \log{[(x_1-x_2)-v_F(t_1-t_2)]}} e^{\frac{1}{2} \log{[(x_1+x_2)+v_F(t_1-t_2)]}}\\
&e^{-A_2\log{[(x_1-x_2)-v_h(t_1-t_2)]}}e^{-B_2\log{[(x_1-x_2)+v_h(t_1-t_2)]}}\\
&e^{-D_2\log{[(x_1+x_2)-v_h(t_1-t_2)]}} e^{-C_2 \log{[(x_1+x_2)+v_h(t_1-t_2)]}}
\end{aligned}
\end{equation*}
\normalsize
Putting $ x = \frac{L}{2} + \epsilon $ and $ x^{'} = - \frac{L}{2} $ so that $ x-x^{'} = L $ and $ x+x^{'} = \epsilon \rightarrow 0 $ is small and also $t_1=t$ and $t_2=0$,
\footnotesize
\begin{equation*}
\begin{aligned}
<T&\psi_R(\frac{L}{2},\sigma,t)\psi_R^{\dagger}(-\frac{L}{2},\sigma,0)>\\
=&\frac{e^{\gamma_1 \log{[L]}}e^{(3+\gamma_2)\log{[L]}}}{2\epsilon} g^{1,1}_{1,-1}(1,1)e^{-E \log{[L-v_Ft ]}}e^{\frac{1}{2} \log{[\epsilon-v_Ft]}}\\
&e^{-A_2\log{[L-v_ht]}}e^{-B_2 \log{[L+v_ht]}}e^{-C_2\log{[\epsilon-v_ht]}} e^{-D_2 \log{[\epsilon+v_ht]}}\\
+&\frac{e^{(3+\gamma_2)\log{[L]}}e^{\gamma_1\log{[L]}} }{2\epsilon}\mbox{  } \mbox{  }g^{1,1}_{1,-1}(1,1)e^{-E \log{[L-v_Ft]}}e^{\frac{1}{2} \log{[\epsilon+v_Ft]}}\\
&e^{-A_2\log{[L-v_ht]}}e^{-B_2\log{[L+v_ht]}}e^{-D_2\log{[\epsilon-v_ht]}} e^{-C_2 \log{[\epsilon+v_ht]}}\\
=&\frac{e^{(3+\gamma_1+\gamma_2)\log{[L]}} }{2\epsilon}\mbox{  }g^{1,1}_{1,-1}(1,1)\mbox{  }e^{-E \log{[L-v_Ft ]}}\mbox{  }e^{\frac{1}{2} \log{[\epsilon-v_Ft]}}\\
&e^{(B_2-A_2)\log{[L-v_ht]}}e^{(D_2-C_2) \log{[\epsilon-v_ht]}}\\
&e^{-B_2 \log{[L^2-(v_ht)^2]}}e^{-D_2\log{[\epsilon^2-(v_ht)^2]}} \\
+&\frac{e^{(3+\gamma_1+\gamma_2)\log{[L]}}}{2\epsilon}\mbox{  }g^{1,1}_{1,-1}(1,1)\mbox{  }e^{-E \log{[L-v_Ft]}}\mbox{  }e^{\frac{1}{2} \log{[\epsilon+v_Ft]}}\\
&e^{(B_2-A_2)\log{[L-v_ht]}}e^{(D_2-C_2) \log{[\epsilon+v_ht]}}\\
&e^{-B_2 \log{[L^2-(v_ht)^2]}}e^{-D_2\log{[\epsilon^2-(v_ht)^2]}} \\
\end{aligned}
\end{equation*}

Now, $D_2-C_2=\frac{1}{2}$\\
\begin{equation*}
\begin{aligned}
<&T\psi_R(\frac{L}{2},\sigma,t)\psi_R^{\dagger}(-\frac{L}{2},\sigma,0)>\\
=&\frac{e^{(3+\gamma_1+\gamma_2)\log{[L]}}}{2\epsilon}g_{1,-1}(1,1)e^{-E \log{[L-v_Ft ]}}e^{ (B_2-A_2)\log{[L-v_ht]}}\\
&e^{\frac{1}{2} \log{[\epsilon-v_Ft]}}e^{\frac{1}{2} \log{[\epsilon-v_ht]}}e^{- B_2\log{[L^2-(v_ht)^2]}}e^{-D_2\log{[\epsilon^2-(v_ht)^2]}} \\
+&\frac{e^{(3+\gamma_1+\gamma_2)\log{[L]}}}{2\epsilon}g_{1,-1}(1,1)e^{-\frac{1}{2} \log{[L-v_Ft]}}e^{(B_2-A_2)\log{[L-v_ht]}}\\
&e^{\frac{1}{2} \log{[\epsilon+v_Ft]}}e^{\frac{1}{2} \log{[\epsilon+v_ht]}}e^{-B_2\log{[L^2-(v_ht)^2]}}e^{-D_2\log{[\epsilon^2-(v_ht)^2]}}\\
\end{aligned}
\end{equation*}

\begin{equation}
\begin{aligned}
\lim_{ \epsilon \rightarrow 0 } \mbox{ }\frac{1}{2\epsilon}\mbox{ }\bigg[&\mbox{ }e^{\frac{1}{2} \log{[\epsilon-v_ht]}}e^{\frac{1}{2} \log{[\epsilon-v_Ft]}}\\
&+e^{\frac{1}{2} \log{[\epsilon+v_ht]}}e^{\frac{1}{2} \log{[\epsilon+v_Ft]}}\bigg] \mbox{ }\to\mbox{ }\frac{v_F+v_h}{2 \sqrt{v_F v_h}}.
\label{epsilon}
\end{aligned}
\end{equation}

\begin{equation*}
\begin{aligned}
<T\psi_R&(\frac{L}{2},\sigma,t)\psi_R^{\dagger}(-\frac{L}{2},\sigma,0)>\\
=&\frac{v_F+v_h}{2 \sqrt{v_F v_h}} \mbox{  }  e^{(3+\gamma_1+\gamma_2)\log{[L]}}  \mbox{  }g_{1,-1}(1,1)\mbox{  } e^{-E \log{[L-v_Ft ]}}\\
&e^{ (B_2-A_2)\log{[L-v_ht]}}e^{- B_2\log{[L^2-(v_ht)^2]}}e^{-D_2\log{[-(v_ht)^2]}}\\
\end{aligned}
\end{equation*}
\normalsize
Since $ G \sim | v_F \int^{ \infty }_{-\infty } dt <\{\psi_R(\frac{L}{2},\sigma,t),\psi_R^{\dagger}(-\frac{L}{2},\sigma,0)\}> |^2 $ it is possible to read off the conductance exponent as follows,
\begin{equation}
G \sim \left( \frac{ L }{ L_{ \omega } }\right)^{2(4+\gamma_1+\gamma_2-A_2-B_2-2D_2-E)}\label{CON1}
\end{equation}
\subsection{Voltage applied across the different poles and on opposite sides of the origin}

In a very similar fashion as described above, the conductance exponent for this case can be obtained as the following.
\begin{equation}
G \sim \left( \frac{ L }{ L_{ \omega } }\right)^{2(4+\gamma_1+\gamma_2-A_3-B_3-2D_3-E)}\label{CON2}
\end{equation}
\subsection{Voltage applied across the different poles but same side of the origin}

Consider the general Green function derived earlier for $ xx^{'} > 0 $ and different poles (Case II). From that it is possible to conclude
($ W = g^{1,2}_{-1,-1}(1,-1)\theta(-x)\theta(-x')+g^{1,2}_{1,1}(1,-1)\theta(x)\theta(x') $),
\footnotesize
\begin{equation*}
\begin{aligned}
<T\psi_R(&x_1,\sigma_1,t_1)\psi_L^{\dagger}(x_2,\sigma_2,t_2)>\\
=&\frac{e^{\gamma_1\log{[2x_1]}}e^{(3+\gamma_2)\log{[2x_2]}}}{2(x_1-x_2)}\mbox{  } \\
&\mbox{   }W\mbox{  }e^{-E \log{[(x_1+x_2)-v_F(t_1-t_2)]}}e^{\frac{1}{2} \log{[(x_1-x_2)-v_F(t_1-t_2)]}}\\
&e^{-A_1\log{[(x_1-x_2)-v_h(t_1-t_2)]}}e^{-B_1 \log{[(x_1-x_2)+v_h(t_1-t_2)]}}\\
&e^{-C_1\log{[(x_1+x_2)-v_h(t_1-t_2)]}} e^{-D_1 \log{[(x_1+x_2)+v_h(t_1-t_2)]}}\\
+&\frac{e^{(3+\gamma_2)\log{[2x_1]}}e^{\gamma_1\log{[2x_2]}} }{2(x_1-x_2)}\mbox{  }\\
&W\mbox{  }e^{-E \log{[(x_1+x_2)-v_F(t_1-t_2)]}} e^{\frac{1}{2} \log{[(x_1-x_2)+v_F(t_1-t_2)]}}\\
&e^{-B_1\log{[(x_1-x_2)-v_h(t_1-t_2)]}}e^{-A_1\log{[(x_1-x_2)+v_h(t_1-t_2)]}}\\
&e^{-C_1\log{[(x_1+x_2)-v_h(t_1-t_2)]}} e^{-D_1 \log{[(x_1+x_2)+v_h(t_1-t_2)]}}\\
\end{aligned}
\end{equation*}
\normalsize
Putting $ x = -\frac{L}{2} + \epsilon $ and $ x^{'} = - \frac{L}{2} $ so that $ x+x^{'} = -L $ and $ x-x^{'} = \epsilon \rightarrow 0 $ is small and also $t_1=t$ and $t_2=0$,
\footnotesize
\begin{equation*}
\begin{aligned}
&<T\psi_R(-\frac{L}{2},\sigma,t)\psi_L^{\dagger}(-\frac{L}{2},\sigma,0)>\\
&=\frac{e^{\gamma_1 \log{[L]}} e^{(3+\gamma_2)\log{[L]}}}{2\epsilon} \mbox{  }W\mbox{  }e^{-E \log{[-L-v_Ft ]}}e^{\frac{1}{2} \log{[\epsilon-v_Ft]}}\\
&e^{-A_1\log{[\epsilon-v_ht]}}e^{-B_1 \log{[\epsilon+v_ht]}}e^{-C_1\log{[-L-v_ht]}} e^{-D_1 \log{[-L+v_ht]}}\\
&+\frac{e^{(3+\gamma_2)\log{[L]}}e^{\gamma_1\log{[L]}} }{2\epsilon}\mbox{  } \mbox{  }W\mbox{  }e^{-E \log{[-L-v_Ft]}}e^{\frac{1}{2} \log{[\epsilon+v_Ft]}}\\
&e^{-A_1\log{[\epsilon+v_ht]}}e^{-B_1\log{[\epsilon-v_ht]}}e^{-C_1\log{[-L-v_ht]}} e^{-D_1 \log{[-L+v_ht]}}\\
\end{aligned}
\end{equation*}%

\begin{equation*}
\begin{aligned}
&<T\psi_R(-\frac{L}{2},\sigma,t)\psi_L^{\dagger}(-\frac{L}{2},\sigma,0)>\\
&=\frac{e^{(3+\gamma_1+\gamma_2)\log{[L]}} }{2\epsilon}\mbox{  } W\mbox{  }e^{-E \log{[-L-v_Ft ]}}e^{\frac{1}{2} \log{[\epsilon-v_ht]}}\\
&\hspace{1cm}e^{(B_1-A_1)\log{[\epsilon-v_ht]}}e^{(C_1-D_1) \log{[-L+v_ht]}}\\
&\hspace{1cm}e^{-B_1 \log{[\epsilon^2-(v_ht)^2]}}e^{-C_1\log{[L^2-(v_ht)^2]}} \\
&+\frac{e^{(3+\gamma_1+\gamma_2)\log{[L]}}}{2\epsilon}\mbox{  }\mbox{  }W\mbox{  }e^{-E \log{[-L-v_Ft]}}e^{\frac{1}{2} \log{[\epsilon+v_ht]}}\\
&\hspace{1cm}e^{(B_1-A_1)\log{[\epsilon+v_ht]}}e^{(C_1-D_1) \log{[-L+v_ht]}}\\
&\hspace{1cm}e^{-B_1 \log{[\epsilon^2-(v_ht)^2]}}e^{-C_1\log{[L^2-(v_ht)^2]}} \\
\end{aligned}
\end{equation*}
Now, $B_1-A_1=\frac{1}{2}$\\
\begin{equation*}
\begin{aligned}
<&T\psi_R(-\frac{L}{2},\sigma,t)\psi_L^{\dagger}(-\frac{L}{2},\sigma,0)>\\
=&\frac{ e^{(3+\gamma_1+\gamma_2)\log{[L]}}}{2\epsilon}W\mbox{  }e^{-E \log{[-L-v_Ft ]}}e^{ (C_1-D_1)\log{[-L+v_ht]}}\\
&e^{\frac{1}{2} \log{[\epsilon-v_Ft]}}e^{\frac{1}{2} \log{[\epsilon-v_ht]}}e^{- B_1\log{[\epsilon^2-(v_ht)^2]}}e^{-C_1\log{[L^2-(v_ht)^2]}} \\
+&\frac{ e^{(3+\gamma_1+\gamma_2)\log{[L]}}}{2\epsilon}W\mbox{  }e^{-\frac{1}{2} \log{[L-v_Ft]}}e^{ (C_1-D_1)\log{[-L+v_ht]}}\\
&e^{\frac{1}{2} \log{[\epsilon+v_Ft]}}e^{\frac{1}{2} \log{[\epsilon+v_ht]}}e^{-B_1\log{[\epsilon^2-(v_ht)^2]}}e^{-C_1\log{[L^2-(v_ht)^2]}}
\end{aligned}
\end{equation*}
Using equation (\ref{epsilon})  for $ \epsilon \rightarrow 0 $,
\begin{equation*}
\begin{aligned}
<T\psi_R&(-\frac{L}{2},\sigma,t)\psi_R^{\dagger}(-\frac{L}{2},\sigma,0)>\\
=&\frac{v_F+v_h}{2 \sqrt{v_F v_h}}
 \mbox{  }  e^{(3+\gamma_1+\gamma_2)\log{[L]}}  \mbox{  }g^{1,2}_{-1,-1}(1,-1)\mbox{  }e^{-E \log{[L-v_Ft ]}}\\
&e^{ (C_1-D_1)\log{[-L+v_ht]}}e^{- C_1\log{[L^2-(v_ht)^2]}}e^{-B_1\log{[\epsilon^2-(v_ht)^2]}} \\
\end{aligned}
\end{equation*}

Since $ G \sim | v_F \int^{ \infty }_{-\infty } dt <\{\psi_R(-\frac{L}{2},\sigma,t),\psi_L^{\dagger}(-\frac{L}{2},\sigma,0)\}>|^2 $ it is possible to read off the conductance exponent as follows,

\begin{equation}
G \sim \left( \frac{ L }{ L_{ \omega } }\right)^{2(4+\gamma_1+\gamma_2-2B_1-C_1-D_1-E)}
\label{CON3}
\end{equation}

\bibliographystyle{apsrev4-1}
\bibliography{ref}
\normalsize

\end{document}